\documentclass[12pt]{article}

\usepackage{graphics,amssymb,amsmath,epsfig,psfrag}
\usepackage{caption, subcaption}
\usepackage[usenames,dvips]{color}
\usepackage{rotating}
\usepackage{hyperref}
\usepackage{float}
\usepackage{enumerate}
\usepackage[numbers, sort&compress]{natbib}
\usepackage{verbatim}

\newcommand{\beq}{\begin{eqnarray}}
\newcommand{\eeq}{\end{eqnarray}}

\newcommand{\eq}[1]{Eq.~(\ref{#1})}
\newcommand{\bib}[1]{Ref.~\cite{#1}}
\newcommand{\fig}[1]{Fig.~\ref{#1}}
\newcommand{\tab}[1]{Table~\ref{#1}}
\newcommand{\sect}[1]{Section~\ref{#1}}

\newcommand{\crn}{\nonumber \\}
\newcommand{\fr}{\frac}

\newcommand{\dd}{{\textrm{d}}}
\newcommand{\hsigma}{{\hat{\sigma}}}
\begin{document}

\begin{titlepage}

\vspace{1cm}

\hfill KA-TP-18-2013

\hfill MPP-2013-221

\hfill SFB/CPP-13-47

\vspace*{1.5cm}

\begin{center}

{\large\bf Massive gauge boson pair production at the LHC:}

\vspace{.1cm}

{\large\bf a next--to--leading order story}

\vspace*{.8cm}

{\large Julien Baglio$^1$, Le Duc Ninh$^{1,2}$ and Marcus M.\ Weber$^3$}

\vspace*{8mm}

$^1$ Institut f\"{u}r Theoretische Physik, Karlsruher Institut f\"{u}r
Technologie,

Wolfgang Gaede Stra{\ss}e 1, Karlsruhe DE-76131, Germany\smallskip

$^2$ Institute of Physics, Vietnam Academy of Science and Technology, 

10 Dao Tan, Ba Dinh, Hanoi, Vietnam\smallskip

$^3$ Max-Planck-Institut f\"{u}r Physik (Werner-Heisenberg-Institut),

D-80805 M\"{u}nchen, Germany

\end{center}

\vspace{1.4cm}

\begin{abstract}
Electroweak gauge boson pair production is one of the most important
Standard Model processes at the LHC, not only because it is a
benchmark process but also by its ability to probe the electroweak
interaction directly. We present full next--to--leading order
predictions for the production cross sections and distributions of
on-shell massive gauge boson pair production in the Standard
Model. This includes the QCD and electroweak corrections. We study the
hierarchy between the different channels when looking at the size of
the QCD gluon-quark induced processes and the electroweak photon-quark
induced processes and provide the first comprehensive explanation of
this hierarchy thanks to analytical leading-logarithmic results. We
also provide a detailed study of the theoretical uncertainties
affecting the total cross section predictions that stem from scale
variation, parton distribution function and $\alpha_s$ errors. We then
compare with the present LHC data.
\end{abstract}

\end{titlepage}

\section{Introduction}

Since the start of LHC operations in 2010 there have been already
several measurements involving electroweak (EW)  gauge bosons,
e.g. the standard candles $W$ and $Z$ production cross
sections~\cite{W-ATLAS, Z-ATLAS, W-CMS, Z-CMS}. In addition to these
standard measurements, the gauge boson pair production mechanisms
provide a very important test of the Standard Model (SM) structure as
it is a way to probe directly the non-abelian structure of EW symmetry
and in particular the trilinear couplings among the $W$, $\gamma$ and
$Z$ bosons. Physics beyond the SM may hide in anomalous
couplings and could be visible in gauge boson pair production at the
LHC. The ATLAS and CMS Collaborations have already provided
measurements and limits on these anomalous gauge boson
couplings~\cite{anomalous-ATLAS-1, ATLAS-WW-7tev,
  anomalous-ATLAS-2, ATLAS-ZZ-7tev, anomalous-ATLAS-3,
  anomalous-ATLAS-4, CMS-ZZ-7tev}. In addition, $WW$ and $ZZ$
production are amongst the most important irreducible backgrounds in
Higgs boson studies, see Ref.~\cite{LHCXS-2} for example. This leads to
the requirement of precise experimental measurements on the one side
and rigorous theoretical understanding on the other side.

The QCD next--to--leading order (NLO) corrections to gauge boson pair
production have been known for a long time~\cite{WZ-QCD-NLO-1,
  WZ-QCD-NLO-2, WW-QCD-NLO, VV-QCD-NLO-1, VV-QCD-NLO-2, VV-QCD-NLO-3},
both for the total cross sections and the differential distributions,
including the decays of the produced gauge
bosons~\cite{QCD-NLO-exclusive, VV-POWHEG}. A full
next--to--next--to--leading order (NNLO) QCD calculation is still
missing even if partial NNLO results have been obtained recently, see
e.g. \bib{WZ-NNLO} for $WZ$ channel and \bib{Dawson:2013lya} for $WW$
channel. Next--to--next--to--leading logarithmic calculations are also
available for $WW$ production~\cite{Kuhn:2011mh}. As for EW
corrections, they have been known for long only in the high energy
approximation~\cite{EW-corrections-early1,   EW-corrections-early2,
  EW-corrections-early3} and only very recently the full NLO EW
corrections have been calculated for diboson
production~\cite{WW-EW-Kuhn, Bierweiler:2012qq, Kuhn-2}, with the
notable exception of photon-quark induced processes.

The goal of this paper is to provide full NLO calculations for the
on-shell $WW$, $WZ$ and $ZZ$ cross sections at the LHC, with both EW
and QCD effects. In particular the EW corrections include photon-quark
initial--state processes that have not yet been considered in the
literature. $WW$ and $ZZ$ cross section calculations also
include the well-known gluon fusion subprocess~\cite{WW-gg-early,
  ZZ-gg-early1, ZZ-gg-early2, WW-gg-recent1, WW-gg-recent2,
  ZZ-gg-recent1} that is formally a NNLO contribution.
Furthermore we include the photon-photon induced process in the
$WW$ cross section calculation. The LO contribution to this
photon-photon induced process has been studied in \bib{WW-EW-Kuhn}. We
also present some distributions for the three different processes,
without any cuts, to study the effect of the separate contributions to
the NLO corrections, especially at high $p_T$. In particular we
present a detailed analysis of the hierarchy that is observed in the
size of the QCD gluon-quark induced corrections and the EW
photon-quark induced corrections. We provide the first comprehensive
explanation of the large differences seen in $ZZ$, $WZ$ and $WW$
channels. There have been attempts to explain them in 
the case of QCD gluon-quark induced corrections~\cite{WZ-QCD-NLO-2,
  WW-QCD-NLO, QCD-NLO-exclusive}. However, to our best knowledge, it
has not been understood why the $WW$ channel is so different from the
$ZZ$ channel~\cite{QCD-NLO-exclusive}. We will show that the
differences are essentially due to non-abelian gauge structure of the
SM, different coupling strengths and parton distribution function
(PDF) effects. Our explanation differs from the one given in
Refs.~\cite{WZ-QCD-NLO-2, WW-QCD-NLO, QCD-NLO-exclusive}. We will also
show that including the photon-quark induced processes is important in the
case of the EW corrections to $WZ$ and $WW$ channels, compensating or
even overcompensating the virtual EW Sudakov effect. In order to
compare with experimental data, we also perform a detailed analysis of
the different sources of uncertainties that affect the theoretical
calculation of the total cross section. This includes the scale
uncertainty that stems from the variation of the renormalization and
the factorization scales, providing an estimate of the missing
higher--order corrections; the uncertainty related to the parton
distribution function (PDF) and the associated error on the
determination of the strong coupling constant $\alpha_s$. The
uncertainties related to the experimental errors on the $W$ and $Z$
masses are found to be negligible. We also study the interplay between
single-top and $WW$ production modes and find that the interference
effects are negligible.

The paper is organized in the following way: in a first section we
present the ingredients of the calculation and in particular the
calculation of the full EW corrections. We then
move to the numerical analysis and present the setup as well as the
distributions for the different processes, using analytical
calculations to explain the hierarchy in the QCD and EW corrections
between $WW$, $WZ$ and $ZZ$ mechanisms. In a third section we
carry out the analysis of the theoretical uncertainties on the total
cross section. We also compare with ATLAS and CMS results. A
short conclusion will then be given. The reader will also find an
appendix where the details of the analytical approximation of the
photon-quark induced process in $W^+Z$ channel are given as an example.

\section{Calculational details\label{sect:cal_details}}

\begin{figure}[h]
  \centering
  \begin{subfigure}[b]{0.5\textwidth}
    \centering
    \includegraphics[scale=0.5]{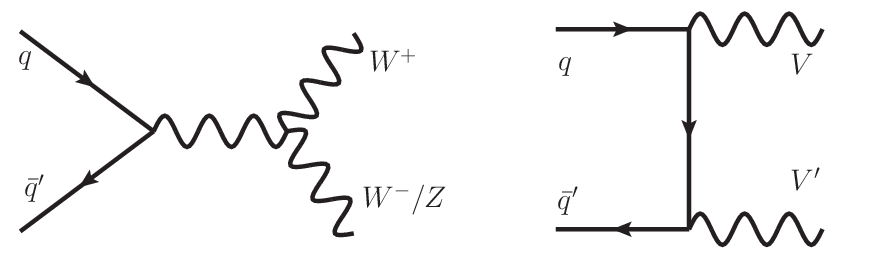}
    \caption{}
    \vspace{5mm}
    \label{fig:qqVV_LO}
  \end{subfigure}
  \begin{subfigure}[b]{0.5\textwidth}
    \centering
    \includegraphics[scale=0.5]{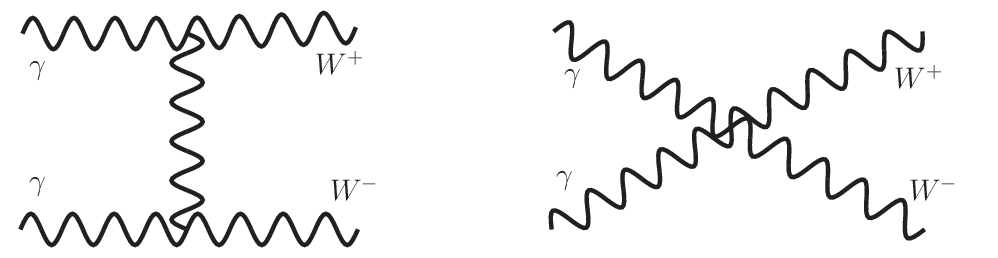}
    \caption{}
    \label{fig:yyVV_LO}
  \end{subfigure}
  \caption{\small Representative tree-level diagrams for $VV^{\prime}$
    production processes.}
  \label{fig:qq_yy_VV_LO}
\end{figure}

\begin{figure}[h]
  \centering
  \includegraphics[width=0.5\textwidth]{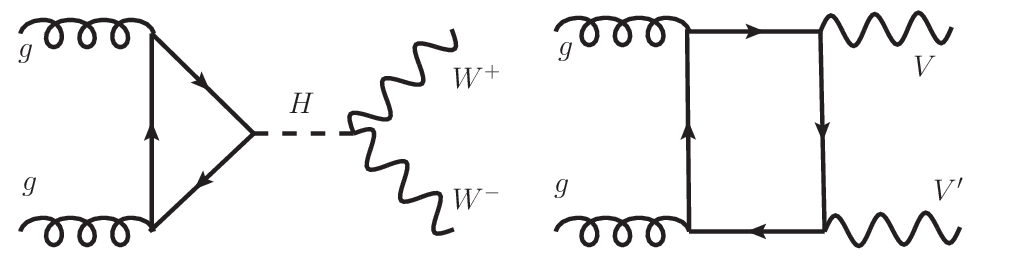}
  \caption{\small Representative diagrams for $VV^{\prime}$ production via
    gluon-gluon fusion.}
  \label{fig:ggVV_LO}
\end{figure}

We consider in this paper the production of two on-shell massive gauge
bosons at the LHC. The contributions from the third-generation quarks
in the initial state are excluded (see the discussion on the $b$-quark
contribution in \sect{bquark}) unless otherwise stated. The main
mechanism to produce two massive gauge bosons is therefore via quark
anti-quark annihilations as shown in \fig{fig:qqVV_LO}. The special
$\gamma\gamma\to W^+W^-$ reaction, which occurs at tree level and
includes the quartic $\gamma\gamma W^+W^-$ coupling as depicted in
\fig{fig:yyVV_LO}, is also taken into account. Even though this
contribution is very small at the total cross section level, it
increases with the invariant mass $M_{WW}$ and is comparable with
opposite sign to the leading EW corrections for the invariant mass
distribution, as shown in \sect{section-WW}. For the case of $W^+W^-$
and $ZZ$, the subdominant one-loop gluon fusion processes, as
displayed in \fig{fig:ggVV_LO} and which are formally NNLO
contributions, are also included in our calculation. These corrections
have been calculated in Refs.~\cite{WW-gg-early, ZZ-gg-early1,
  ZZ-gg-early2, WW-gg-recent1, WW-gg-recent2, ZZ-gg-recent1}. We
re-calculate them here for the sake of completeness.
The leading order (LO) hadronic cross section is defined as
\beq
\sigma_\text{LO}= \int \dd x_1\dd x_2[\bar{q}_\text{LO}(x_1,
\mu_F)q^{\prime}_\text{LO}(x_2, \mu_F)\hsigma^{\bar{q}q^{\prime}\to
  VV^{\prime}}_\text{LO}+(1\leftrightarrow 2)],
\label{xsection_LO}
\eeq
where $q$ and $\bar{q}$ are the parton distribution functions (PDF) of
the first and second generation quarks in the proton at the momentum
fraction $x$ and factorization scale $\mu_F$, and
$\hsigma^{\bar{q}q^{\prime}\to VV^{\prime}}$ the LO partonic cross
section. The photon-photon and gluon-gluon contributions read
\beq
\sigma^{\gamma\gamma}_\text{LO} &=& \int \dd x_1\dd
x_2[\gamma_\text{LO}(x_1, \mu_F)\gamma_\text{LO}(x_2, \mu_F)
\hsigma_\text{LO}^{\gamma\gamma\to VV^{\prime}}],\crn
\sigma^{gg} &=& \int \dd x_1\dd x_2[g(x_1, \mu_F)g(x_2,
\mu_F)\hsigma^{gg\to VV^{\prime}}]\, ,
\label{xsection_gam_glu_LO}
\eeq
and are understood as corrections over the LO hadronic cross section
of \eq{xsection_LO}.

In the following, we sketch the main points of our NLO calculation. We
will define various sub-corrections at NLO, namely the QCD virtual,
gluon-quark radiated and gluon-quark induced corrections for the QCD case and the
EW virtual, photon-quark radiated and photon-quark induced corrections for the EW
case. These sub-corrections are ultraviolet (UV) and infrared (IR)
finite, but are dependent on the regularization scheme. The final
results, {\it i.e.} the sum of those sub-corrections, are of course
regularization-scheme independent. The separation will help to better
understand the QCD and EW corrections.

\subsection{NLO QCD corrections\label{sect:NLOQCD}}

The NLO QCD contribution includes the loop corrections with one gluon
in the loops and the real emission corrections with one additional
parton in the final state. We classify the real emission contribution
into two groups: gluon-quark radiated processes $\bar{q}q^{\prime}\rightarrow
VV^{\prime}g$, where the gluon is radiated from an initial
(anti-)quark and gluon-quark induced processes $qg\rightarrow
VV^{\prime}q^{\prime}$, which are related to the gluon-quark radiated ones
via crossing symmetry. Both the virtual and real corrections are
separately IR divergent. These divergences cancel in the sum for
infrared-safe observables such as the total cross section and
kinematic distributions of massive gauge bosons.
The dimensional regularization (DR) method~\cite{DR-method} is
used to regularize the infrared divergences unless otherwise
stated. Moreover, we apply the Catani-Seymour dipole subtraction
algorithm~\cite{dipole-substraction} to combine the virtual and the
real contributions. We use the same notations as in
\bib{dipole-substraction} and define the various NLO QCD corrections as
follows,
\beq
\sigma_\text{QCD-virt} &=& \int \dd x_1\dd x_2[\bar{q}_\text{NLO}(x_1,
\mu_F)q^{\prime}_\text{NLO}(x_2, \mu_F)\hsigma^{\bar{q}q^{\prime}\to
  VV^{\prime}}_\text{QCD-virt}+(1\leftrightarrow 2)],\crn
\hsigma^{\bar{q}q^{\prime}\to VV^{\prime}}_\text{QCD-virt} &=&
\hsigma^{\bar{q}q^{\prime}\to VV^{\prime}}_\text{QCD-loop} +
\hsigma^{\bar{q}q^{\prime}\to VV^{\prime}}_\text{QCD-I},\,
\label{xsection_virt_qcd}
\eeq
where $\hsigma^{\bar{q}q^{\prime}\to VV^{\prime}}_\text{QCD-loop}$
includes only loop diagrams and $\hsigma^{\bar{q}q^{\prime}\to
  VV^{\prime}}_\text{QCD-I}$ is the I-operator contribution as defined
in \bib{dipole-substraction}. It is noted that
$\hsigma^{\bar{q}q^{\prime}\to VV^{\prime}}_\text{QCD-virt}$ is UV and
IR finite. The gluon-quark radiated and gluon-quark induced contributions read
\beq
\sigma_\text{g-rad} &=& \int \dd x_1\dd x_2[\bar{q}_\text{NLO}(x_1,
\mu_F)q^{\prime}_\text{NLO}(x_2, \mu_F)
\left(\hsigma^{\bar{q}q^{\prime}\rightarrow VV^{\prime}g} -
  \hsigma^{\bar{q}q^{\prime}\rightarrow
    VV^{\prime}}_\text{QCD-I}\right)+(1\leftrightarrow 2)],\crn
\sigma_\text{g-ind} &=& \int \dd x_1\dd x_2[q_\text{NLO}(x_1,
\mu_F)g_\text{NLO}(x_2, \mu_F)
\hsigma^{qg\rightarrow VV^{\prime}q^{\prime}} + (1\leftrightarrow
2)].
\label{xsection_real_qcd}
\eeq
These contributions are also IR finite because the collinear
divergences occurring at the partonic level are absorbed into the
quark PDFs.

\subsection{NLO EW corrections\label{sect:NLOEW}}

\begin{figure}[h]
\begin{center}
\includegraphics[scale=0.5]{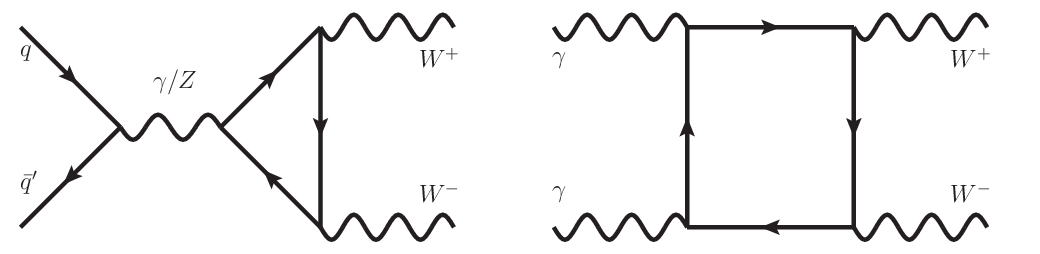}
\caption{\small Selected fermion-loop diagrams with $\gamma_5$.}
\label{fig:ppWW_fermion_loop}
\end{center}
\end{figure}

The NLO EW corrections are also divided into similar sub-contributions
as in the QCD case, but there are some important differences. The loop
corrections contain UV divergences. They are regularized using the DR
method~\cite{DR-method, DR-axial} and by the renormalization of the
relevant EW parameters, namely $M_W$, $M_Z$ and the fine-structure
constant $\alpha$. The presence of fermion loops with $\gamma_5$, as
shown in \fig{fig:ppWW_fermion_loop},
requires that all lepton and quark contributions must be included to
cancel the anomaly.
For the process $\gamma\gamma \to W^+ W^-$ we use
the on-shell scheme (see e.g. \bib{on-shell-scheme}) where
$\alpha(q^2)$  is defined in the Thomson limit at $q^2 \to 0$. This is
justified because the LO amplitudes involve real photons, hence the
running of $\alpha$ is absent. We note that the virtual and
soft-photonic corrections to this process have been calculated in
\bib{photon-corrections}. For $\bar{q}q^{\prime}\to VV^\prime$ processes,
using $\alpha(0)$ as an input parameter induces large corrections of
the form $\log(Q^2/m_f^2)$ with $Q$ being a typical hard energy scale
and $m_f$ the light fermion masses in the loop contribution. Those
corrections can be absorbed into the running of $\alpha$ using
$\alpha(M_Z^2)$ or using the $G_\mu$-scheme with
$\alpha_{G_\mu}=\sqrt{2}G_\mu M_W^2(1-M_W^2/M_Z^2)/\pi$ as an input
parameter. We choose the latter and hence the calculation is
consistently done by fixing the renormalization constant of the
electric charge as
\begin{align}
\delta Z_{e}^{G_\mu} = \delta Z_{e}^{\alpha(0)} - \fr{1}{2}(\Delta
r)_\text{1-loop},
\label{eq:charge-renormalization}
\end{align}
where $(\Delta r)_\text{1-loop}$ is defined in
\cite{muon-correction-1, muon-correction-2}. An advantage of this
framework is that the final results are independent of the light
fermion masses. The above discussion also makes it clear that one
should use the coupling $\alpha(0)$ for vertices with a real photon
directly attached to them. Therefore, the NLO EW corrections to
$\bar{q}q^{\prime}\to VV^\prime$ processes are proportional to
$\alpha_{G_\mu}^2\alpha(0)$, while it is $\alpha(0)^3$ for
$\gamma\gamma \to W^+ W^-$. We now take into account the real
corrections and combine them with the virtual ones using the notation
of \bib{photon-radiation}. We use by default the mass regularization (MR)
method, {\it i.e.} introducing a common mass regulator for the light
fermions (all but the top quark) and a fictitious photon mass, to
regularize IR divergences. For $\gamma\gamma \to W^+ W^-$ process, we
have
\beq
\sigma^{\gamma\gamma}_\text{NLO} &=& \int \dd x_1\dd
x_2[\gamma_\text{NLO}(x_1, \mu_F)\gamma_\text{NLO}(x_2,
\mu_F)\hsigma_\text{NLO}^{\gamma\gamma\to VV^{\prime}}],\crn
\hsigma_\text{NLO}^{\gamma\gamma\to VV^{\prime}} &=&
\hsigma_\text{LO}^{\gamma\gamma\to VV^{\prime}} +
\hsigma_\text{EW-loop}^{\gamma\gamma\to VV^{\prime}} +
\hsigma^{\gamma\gamma\to VV^{\prime}\gamma}.
\label{xsection_gam_NLO}
\eeq
For the $\bar{q}q^{\prime}\to VV^\prime$ processes, the correction
$\sigma_\text{EW-virt}$ is, similarly to the QCD case, given as in
\eq{xsection_virt_qcd}, but the I-operator contribution
$\hsigma^{\bar{q}q^{\prime}\to VV^{\prime}}_\text{EW-I}$ is now
defined as the endpoint contribution of \bib{photon-radiation}. The
photon-quark radiated and photon-quark induced contributions read
\beq
\sigma_{\gamma\text{-rad}} &=& \int \dd x_1\dd
x_2[\bar{q}_\text{NLO}(x_1, \mu_F)q^{\prime}_\text{NLO}(x_2, \mu_F)
\left(\hsigma^{\bar{q}q^{\prime}\rightarrow VV^{\prime}\gamma} -
  \hsigma^{\bar{q}q^{\prime}\rightarrow
    VV^{\prime}}_\text{EW-I}\right)+(1\leftrightarrow 2)],\crn
\sigma_{\gamma\text{-ind}} &=& \int \dd x_1\dd x_2[q_\text{NLO}(x_1,
\mu_F)\gamma_\text{NLO}(x_2, \mu_F)
\hsigma^{q\gamma\rightarrow VV^{\prime}q^{\prime}} + (1\leftrightarrow
2)].
\label{xsection_real_ew}
\eeq
For EW corrections, we use $f_\text{NLO}(x, \mu_F) = f_\text{LO}(x,
\mu_F)$ for $f=q,\bar{q},\gamma$ as given by the current only PDF set
that exists including the photon PDF\footnote{The NNPDF
Collaboration has recently started a QED study of parton
functions~\cite{NNPDF-QED}. A new QED PDF has just been released in
the LHAPDF framework.}, namely
{\texttt{MRST2004QED}}~\cite{PDF-QED}. Moreover, the collinear
divergences occurring at the partonic level in the photon-quark radiated and
photon-quark induced contributions are absorbed into the (anti-)quark and
photon PDFs using the DIS factorization scheme as done in
\bib{collinear-photons}. We note that the virtual and photon-quark radiated
contributions have been calculated in Refs.~\cite{Bierweiler:2012qq,
  WW-EW-Kuhn, Kuhn-2}, but the photon-quark induced processes were not
considered and the $\gamma\gamma$ contribution was considered at LO
only.

In addition to the MR scheme, we have also implemented the DR scheme
to deal with IR divergences for $\bar{q}q^\prime \to VV^\prime$
processes. For QCD corrections the DR method is explained in detail in
\bib{Bredenstein:2008zb}. For EW corrections the same procedure
holds. This is because, as in the QCD case, the rational terms of the
IR origin come only from the collinear single pole in the
wave-function corrections of massless particles. The soft divergences
related to the $W^+W^-\gamma$ vertex do not introduce any rational
term. Moreover, by using the $\alpha_{G_\mu}$ scheme, the results are
independent of the light fermion masses. These masses can therefore be
set to zero. We have checked that, by using the same subtraction term
in the $2 \to 3$ contribution, the sum of the $\text{I}$-operator and
the $\text{PK}$-operator (as defined in \bib{dipole-substraction} for
DR and being the convolution part in \bib{photon-radiation} for MR)
contributions is in agreement within statistical errors for the MR and
the DR methods.

The aforementioned method has been implemented in different computer
codes, using {\texttt{FORTRAN77}} and {\texttt{C++}} programming
languages. The helicity amplitudes are generated using
{\texttt{FeynArts-3.4}}~\cite{FeynArts} and
{\texttt{FormCalc-6.0}}~\cite{FormCalc} as well as
{\texttt{HELAS}}~\cite{HELAS, MadGraph}. The scalar and tensor
one-loop integrals are evaluated with the in-house library
{\texttt{LoopInts}}, which agrees with
{\texttt{LoopTools}} program~\cite{FormCalc, Form}. In these codes,
the tensor integrals are recursively reduced to scalar integrals using
the Passarino-Veltman algorithm~\cite{PaVeReduction} and the scalar
integrals are calculated as in Refs.~\cite{scalar-integral-1,
  scalar-Dittmaier, scalar-Nhung, scalar-Denner}. 
Moreover, the real corrections have been checked
by comparing the results of the dipole-subtraction method with those
of the phase-space slicing method~\cite{slicing-method}.

\section{Differential cross sections\label{section-3}}

\subsection{Parameter setup\label{section-parameters}}

Our default set of input parameters is
\begin{align}
G_{\mu} = 1.16637\times 10^{-5} \text{ GeV}^{-2}, \, &
M_W=80.385 \text{ GeV}, \, M_Z = 91.1876 \text{ GeV}, \nonumber\\ 
 M_t = 173.5 \text{ GeV}, \, & M_H=125 \text{ GeV},
\label{param-setup}
\end{align}
taken from Refs.~\cite{RPP-2012,ATLAS-Higgs,CMS-Higgs}. The CKM matrix
is assumed to be diagonal. The effect is at most $-0.7\% (-1\%)$ on
the total $W^+ Z (W^- Z)$ cross section. The masses of
the leptons and the light quarks, {\it i.e.} all but the top mass, are
approximated as zero. This is justified because our results are
insensitive to those small masses. As argued in \sect{sect:NLOEW}, the
NLO EW corrections to $\bar{q}q^{\prime}\to VV^\prime$ processes are
proportional to $\alpha_{G_{\mu}}^2\alpha(0)$ where
$\alpha_{G_\mu}=\sqrt{2}G_\mu M_W^2(1-M_W^2/M_Z^2)/\pi$ (which can be
understood as $\alpha(M_W^2)$) and $\alpha(0)=1/137.036$ is used as an
input parameter. This is because the relation
$\alpha_{G_\mu}=\alpha(0)(1+\Delta r)$ depending on the hadronic
contribution to the photon self-energy at low energy is not reliable
and hence we do not use it to calculate $\alpha(0)$. We only list in
this section the central values for the parameters, the uncertainties
that affect them will be described in \sect{section-4}. For the
{\texttt{MRST2004QED}} set of PDFs we use
$\alpha^\text{NLO}_s(M_Z^2)=0.1190$ and the NLO running with five
flavors for all subprocesses. When switching to modern PDF sets such
as {\texttt{MSTW2008}}~\cite{PDF-MSTW}, the difference is
$\alpha_s^\text{NLO}(M_Z^2) = 0.12018$ and
$\alpha_s^\text{NNLO}(M_Z^2)=0.11707$ (in this example). The NNLO
value (and running) of the strong coupling constant is used for $gg\to
W^+W^-,ZZ$ subprocesses. Otherwise the strong coupling constant gets
its NLO value and the running is also evaluated at NLO. We also define
the following central scale for the process $pp\to VV^\prime$,
\begin{align}
\mu_R = \mu_F = \mu_0 = M_{V}+M_{V^\prime}.
\end{align}
This choice is justified because, as we will see, the bulk of the
total cross section comes from the low energy regime. In the
following, we will study the distributions using only the
{\texttt{MRST2004QED}} PDF set. We apply no cuts at the level of the
$W^\pm$ and $Z$, since these will decay. Analytical results for
leading-logarithmic corrections arising from the QCD gluon-quark induced
processes and the EW photon-quark induced processes will also be
presented. Their proofs are given in the Appendix.

\subsection{\texorpdfstring{$ZZ$}{ZZ} distributions\label{section-ZZ}}

We start the discussion of the distributions with the $ZZ$ process. In
this subsection and the following we display the transverse momentum
of one of the gauge bosons, the invariant mass and the rapidity
distributions of the gauge boson pair. The considered center-of-mass
(c.m.) energy will be $\sqrt{s}=14$ TeV. We use $\Delta K_X =
d\sigma_X / d\sigma_{\rm LO}$ to describe the various corrections when
$\Delta K_{\rm NLO}>1$, otherwise we will discuss in terms of
percentage corrections. We use the {\tt MRST2004QED} PDF set both at
LO and NLO.

\begin{figure}[ht!]
\begin{center}
\includegraphics[scale=0.7]{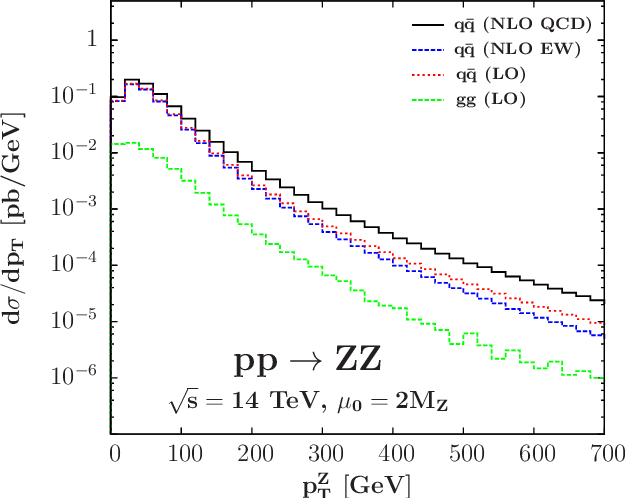}
\hspace{10mm}
\includegraphics[scale=0.7]{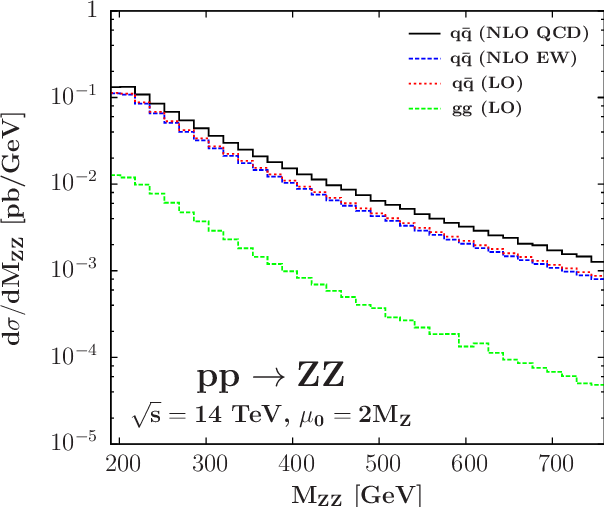}
\end{center}
\it{\vspace{-6mm}\caption{\small $Z$ transverse momentum $p_T$ (in
    GeV) distribution (left) and $Z$ pair invariant mass $M_{Z\!Z}$ (in
    GeV) distribution (right) of $pp\to ZZ$ cross section at the LHC
    (in pb/GeV), including NLO QCD and EW corrections calculated with
    {\texttt{MRST2004QED}} PDF set and with the input parameters described in
    Section~\protect\ref{section-parameters}. In the $M_{ZZ}$
    distribution the $q\bar{q}$ LO (dotted red) and $q\bar{q}$ NLO EW
    (dashed blue) curves nearly coincide.\label{ZZ-all-distribution}}}
\end{figure}

The first distributions that we display in \fig{ZZ-all-distribution}
are the transverse momentum $p_T^{Z}$ of one $Z$ boson as well as the
differential distribution of the invariant mass $M_{Z\!Z}$ of the $Z$
boson pair, calculated at LO and NLO including QCD and EW
corrections. In practice, since the two $Z$ bosons in the final state
are identical, the binning of the $p_T^{Z}$ histogram is done by
selecting both particles and rescaling the final result by symmetry
factor $1/2$. The $gg$ contribution is also separately shown.

\begin{figure}[ht!]
\begin{center}
\includegraphics[scale=0.7]{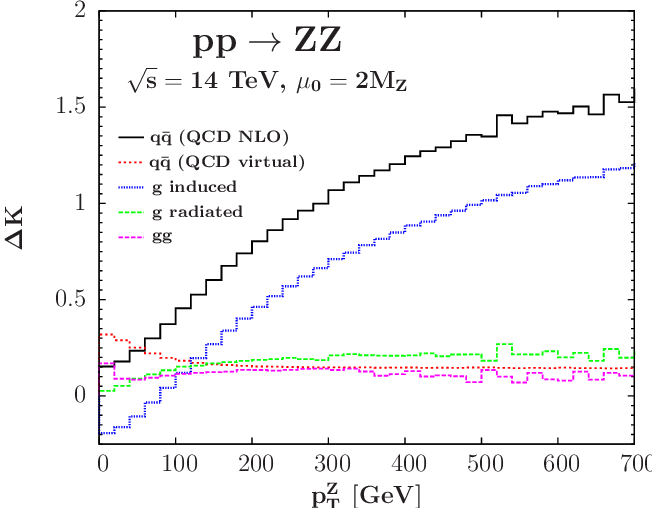}
\hspace{10mm}
\includegraphics[scale=0.7]{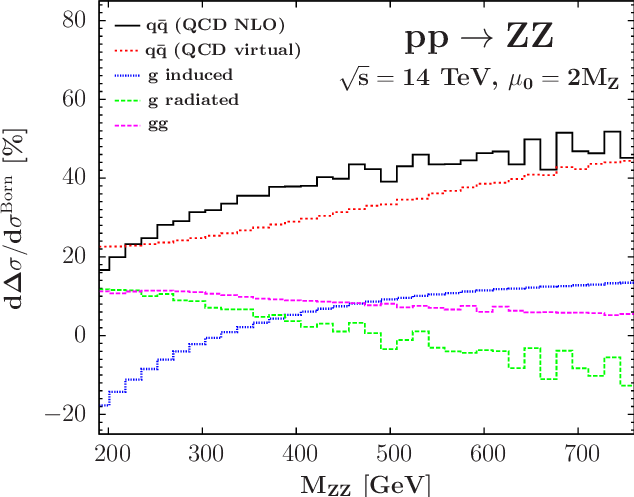}
\end{center}
\it{\vspace{-6mm}\caption{\small $Z$ transverse momentum $p_T$ (in TeV)
    distribution (left, using $\Delta K$) and $Z$ pair invariant mass $M_{Z\!Z}$ (in GeV)
    distribution (right, in $\%$) of the NLO QCD corrections to $pp\to ZZ$ cross
    section at the LHC, calculated with {\texttt{MRST2004QED}} PDF set
    and with the input parameters described in
    Section~\protect\ref{section-parameters}. \label{ZZ-QCD-distribution}}}
\end{figure}

\begin{figure}[ht!]
\begin{center}
\includegraphics[scale=0.7]{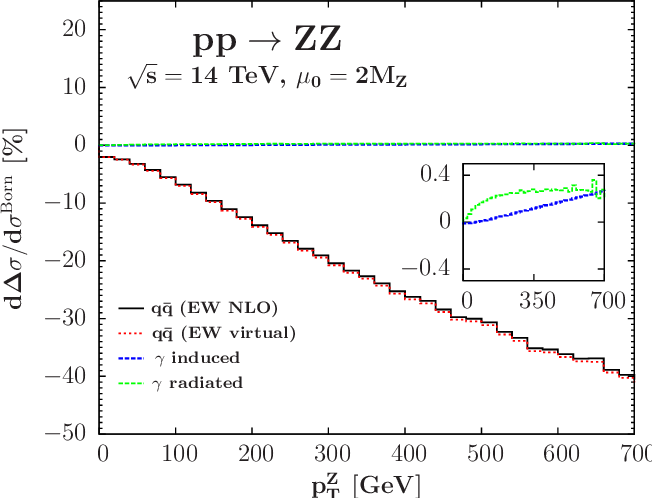}
\hspace{10mm}
\includegraphics[scale=0.7]{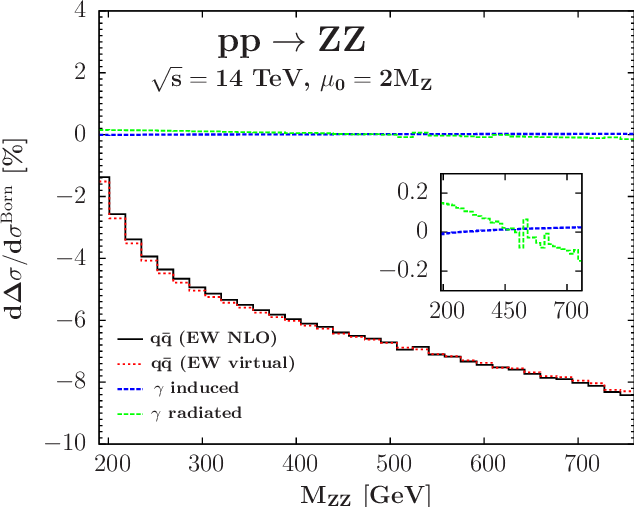}
\end{center}
\it{\vspace{-6mm}\caption{\small Same as
    \protect\fig{ZZ-QCD-distribution} but for EW
    corrections (in $\%$). The $q\bar{q}$ EW NLO and $q\bar{q}$ EW virtual
    curves nearly coincide as well as the photon-quark induced and
    photon radiated curves.\label{ZZ-EW-distribution}}}
\end{figure}

To analyze in more details we display in \fig{ZZ-QCD-distribution} and
\fig{ZZ-EW-distribution} the $p_T^{Z}$ and $M_{Z\!Z}$ distributions of
the QCD and EW corrections, relative to the LO $q\bar q$ results, for
the various sub-corrections defined in \sect{sect:cal_details}. For
the $p_T^{Z}$ distributions, the QCD corrections have the expected
behavior known from the previous studies~\cite{WZ-QCD-NLO-2,
  WW-QCD-NLO}, namely $qg \to ZZ q$ processes dominate entirely the
NLO QCD corrections of the $q\bar q$ subprocesses, up to $\Delta
K =1.5$ at $p_T^{Z}=700~\text{GeV}$. It is
important to note that the $d\bar{d}$ contribution is about $1.3$
times larger than the $u\bar{u}$ one for a large range of $p_T^{Z}$
about $700~\text{GeV}$ at LO because of larger couplings, while the
$dg \to ZZ d$ and $ug \to ZZ u$ contributions are about the same. The
$\bar{q}g \to ZZ \bar{q}$ processes are much smaller. This large
correction can be explained as follows. At large $p_T^{Z}$, both $Z$
bosons are hard with the same transverse momentum at LO. For the
process $qg \to ZZ q$, the dominant mechanism is first to produce one
$Z$ and a quark with large transverse momentum, and then the hard
quark radiates a soft $Z$. For $p_T^{Z} \gg M_Z$ we have, see the
Appendix for more details,
\begin{align}
\fr{d\sigma^{qg \to ZZ q}}{d\sigma_\text{LO}^{\bar{q}q \to ZZ}}
&= \fr{c_{L,q}^2d\sigma^{qg \to Z q}_L + c_{R,q}^2d\sigma^{qg \to Z q}_R}
{4d\sigma_\text{LO}^{\bar{q}q \to
    ZZ}}\fr{\alpha}{2\pi}\log^2\left[\fr{(p_T^{Z})^2}{M_Z^2}\right]
\crn &=c^q_{ZZ} \fr{d\sigma^{qg \to Z q}_L}{d\sigma_\text{LO}^{\bar{q}q
    \to ZZ}}
\fr{\alpha}{2\pi}\log^2\left[\fr{(p_T^{Z})^2}{M_Z^2}\right],
\label{eq_log2_zz}
\end{align}
where $c_{L,f}=(I_3^f - \sin^2\theta_W
Q_f)/(\sin\theta_W\cos\theta_W)$ and $c_{R,f}=-Q_f
\sin\theta_W/\cos\theta_W$ with $I_3^f = \pm 1/2$ being the third
component of the weak isospin and $Q_f$ being the electric charge of
the fermion $f$, and $c^q_{ZZ}=(c_{L,q}^4+c_{R,q}^4)/(4c_{L,q}^2)$
about $0.18$ for up quarks and $0.26$ for down quarks. A sum over $u$
and $d$ flavors is implicitly assumed in \eq{eq_log2_zz}. At $p_T =
700$ GeV we have
\begin{align}
\fr{\alpha}{2\pi}\log^2\left[\fr{(p_T)^2}{M_Z^2}\right] \approx 0.020,
\quad
\fr{\alpha}{2\pi}\log^2\left[\fr{(p_T)^2}{M_W^2}\right] \approx
0.023.
\end{align}
The first ratio on the right-hand side of \eq{eq_log2_zz} is large
because the denominator is an EW process while the numerator is
proportional to $\alpha_s(\mu_0)$ and involves the large gluon PDF.
Numerically, we get $\Delta K_{q g} = 1.51$ ($1.20$) for
leading-logarithmic approximation (full calculation) at $p_T^{Z}=700$
GeV.

In the invariant mass distribution the gluon-quark induced and the
gluon-quark radiated processes compensate each other and the full QCD
corrections are dominated by the virtual corrections. The gluon fusion
mechanism amounts to $\sim +10\%$ in the $M_{Z\!Z}$ distribution and
is negligible in the transverse momentum distribution.

Turning to EW corrections in \fig{ZZ-EW-distribution}, we observe that
the effect of the virtual Sudakov logarithms $\alpha
\log^2[(p_T^{Z})^2/M_W^2]$ is clearly visible in
\fig{ZZ-EW-distribution} (left) where the dotted red curve is the
$p_T^{Z}$ distribution of the virtual EW corrections, and we have for
example a $-40\%$ correction at $p_T^{Z}=700$ GeV. The photon-quark induced
processes have similar behaviors as the gluon-quark induced processes,
except that there are now the photon PDF and the EW $\alpha(0)$
coupling, instead of the gluon PDF and $\alpha_s$, in the right-hand
side of \eq{eq_log2_zz}. For $p_T^{Z} \gg M_Z$ we have
\begin{align}
d\sigma^{q\gamma \to ZZ q}
&=c^q_{ZZ} d\sigma^{q\gamma \to Z q}_L
\fr{\alpha}{2\pi}\log^2\left[\fr{(p_T^{Z})^2}{M_Z^2}\right].
\label{eq_log2_zz_EW}
\end{align}
This explains why the correction is negligible.
Numerically, we get $0.3\%$ ($0.3\%$) for
leading-logarithmic approximation (full calculation) at $p_T^{Z}=700$ GeV.
The same effect is visible in the invariant mass distribution, where the total
EW corrections are dominated by the virtual EW contribution and
drop to $-8\%$ at $M_{Z\!Z} = 700$ GeV.

\begin{figure}[ht!]
\begin{center}
\includegraphics[scale=0.7]{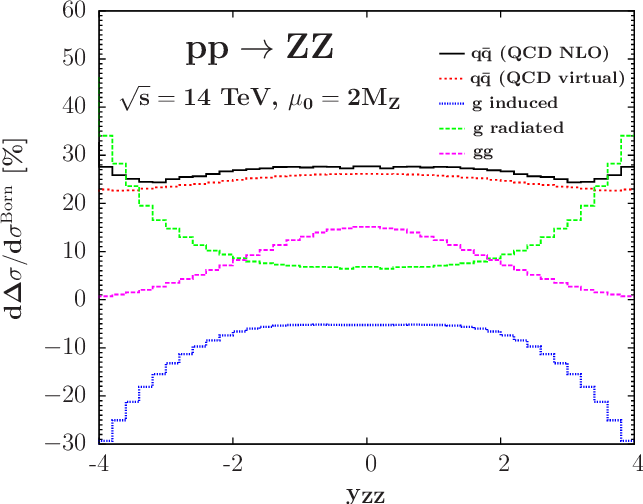}
\hspace{6mm}
\includegraphics[scale=0.7]{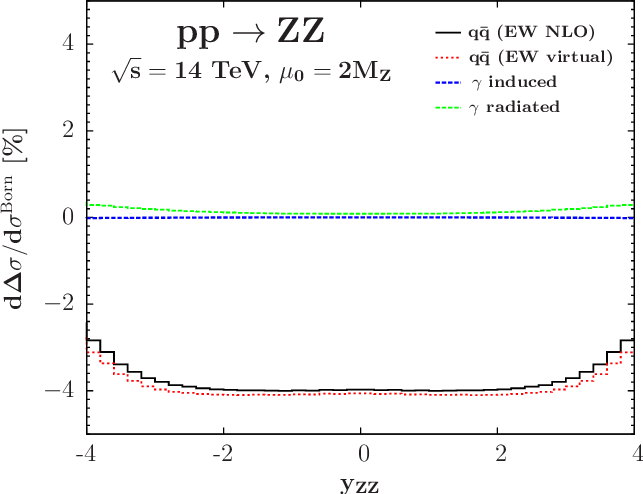}

\vspace{3mm}
\includegraphics[scale=0.7]{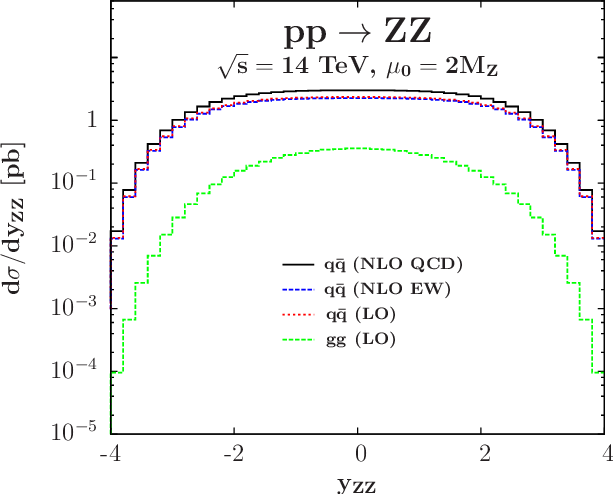}
\end{center}
\it{\vspace{-6mm}\caption{\small $Z$ pair rapidity $y_{Z\!Z}$
    distribution of the NLO QCD corrections and NLO EW corrections
    (upper left and right, respectively, in $\%$), as well as the
    distribution of the $pp\to ZZ$ cross section (lower, in pb) at
    the LHC, calculated with {\texttt{MRST2004QED}} PDF set and with the input
    parameters described in
    Section~\protect\ref{section-parameters}. The $q\bar{q}$ LO
    (dotted red) and $q\bar{q}$ NLO EW (dashed blue) curves nearly
    coincide in the lower panel.\label{ZZ-rapidity-distributions}}}
\end{figure}

We also display the rapidity distribution of the $Z$ pair in
\fig{ZZ-rapidity-distributions}, that gives the trend of
$\log(x_1/x_2)$ momentum fraction of the incoming partons. The impact
of the gluon fusion channel is much smaller than the NLO QCD
corrections on $q\bar q$ sub-channels as seen in the upper left
panel. The latter corrections amount to a $\sim +26\%$ increase. The
upper right panel of \fig{ZZ-rapidity-distributions} displays the
rapidity distribution of the EW corrections and the same effect seen
in the transverse momentum and invariant mass distributions is clearly
visible: the photon-quark induced and photon radiated processes do not contribute
at all and all the EW corrections are driven by the virtual
corrections. The shape is then quadratic with a flat minimum of $\sim
-4\%$ for $|y_{Z\!Z}| \leq 3.5$. This result is consistent with 
the one of \bib{Kuhn-2}.

\vspace{-1mm}

\vspace{-1mm}

\subsection{\texorpdfstring{$WZ$}{WZ} distributions\label{section-WZ}}

\begin{figure}[ht!]
\begin{center}
\includegraphics[scale=0.7]{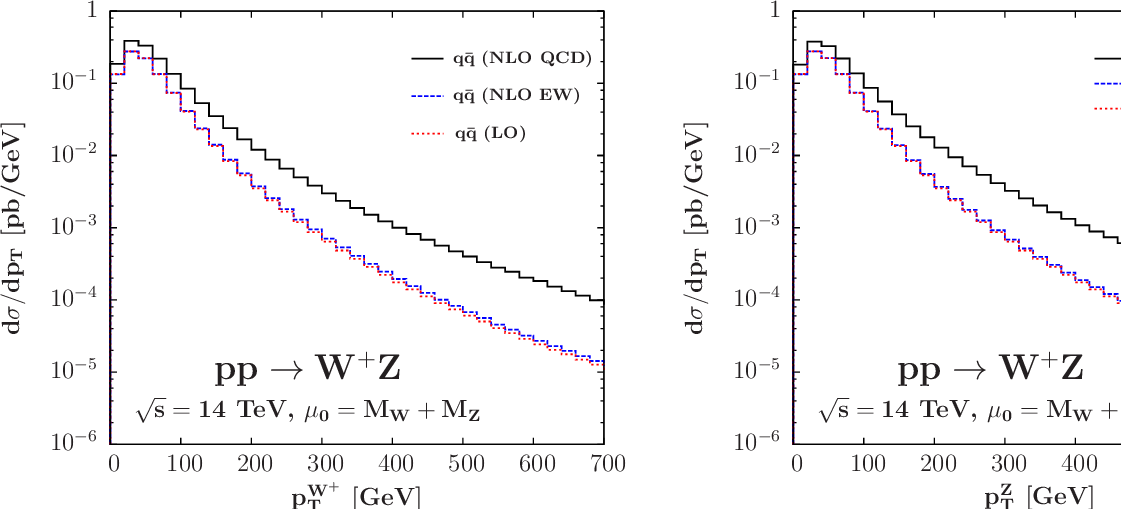}

\vspace{2mm}
\includegraphics[scale=0.7]{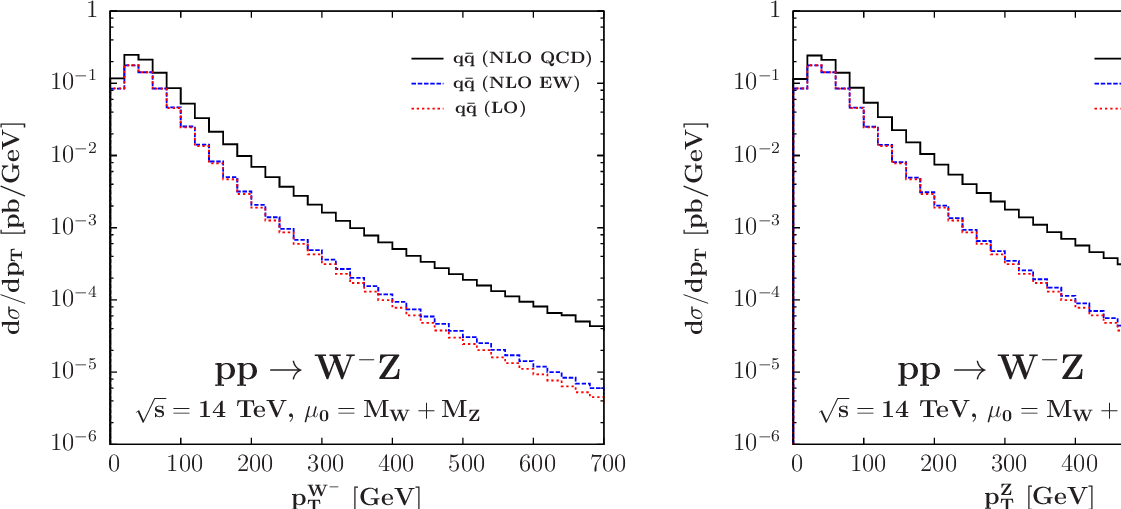}
\end{center}
\it{\vspace{-6mm}\caption{\small $W$ and $Z$ transverse momentum $p_T$
    (in GeV) distributions of $pp\to W^+Z$ (upper left and right
    respectively) and $pp\to W^-Z$ (lower left and right respectively)
    cross sections at the LHC (in pb/GeV), including NLO QCD and EW
    corrections calculated with {\texttt{MRST2004QED}} PDF set and with the input
    parameters described in
    Section~\protect\ref{section-parameters}. The $q\bar{q}$ LO
    (dotted red) and $q\bar{q}$ NLO EW (dashed blue) curves nearly
    coincide.\label{WZ-all-pT-distribution}}}
\end{figure}

We now present the same distributions as above but for the $W^+Z$ and
$W^-Z$ channels. We start with the transverse momentum distributions
of the $W^{\pm}$ boson and of the $Z$ boson displayed in
\fig{WZ-all-pT-distribution}, including NLO QCD and EW
corrections. Both $W^+Z$ and $W^-Z$ channels are depicted and one can
see the same trend in both channels for both $W^{\pm}$ and $Z$ transverse
momenta. The main difference between $W^+Z$ and $W^-Z$ channels comes
from the EW corrections at high transverse momentum: in the first case
there are basically no EW corrections (upper panels)  while in the
latter case the EW corrections increase the differential cross section
by a limited amount (lower panels).

\begin{figure}[ht!]
\begin{center}
\includegraphics[scale=0.7]{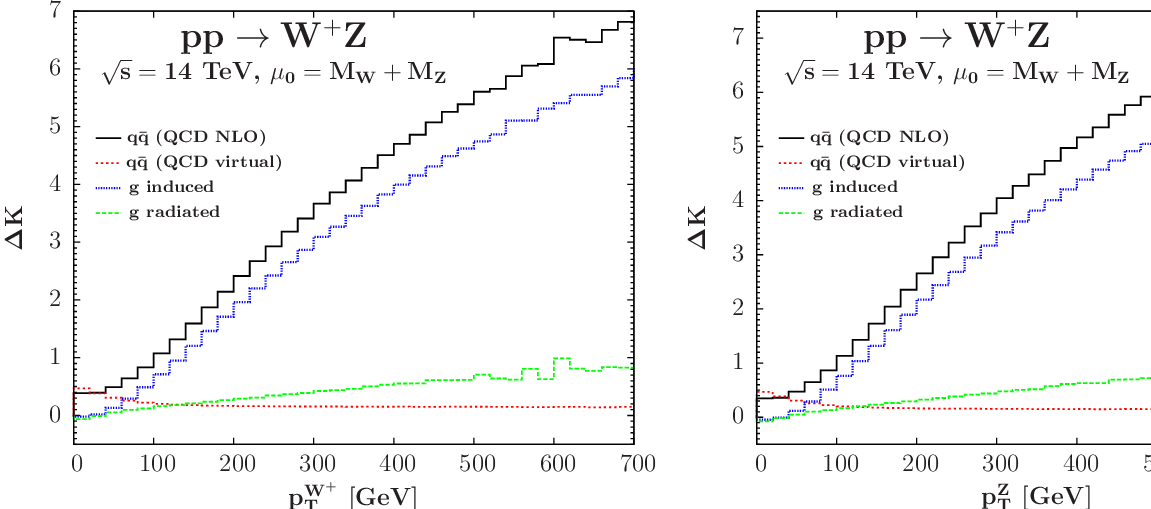}

\vspace{2mm}
\includegraphics[scale=0.7]{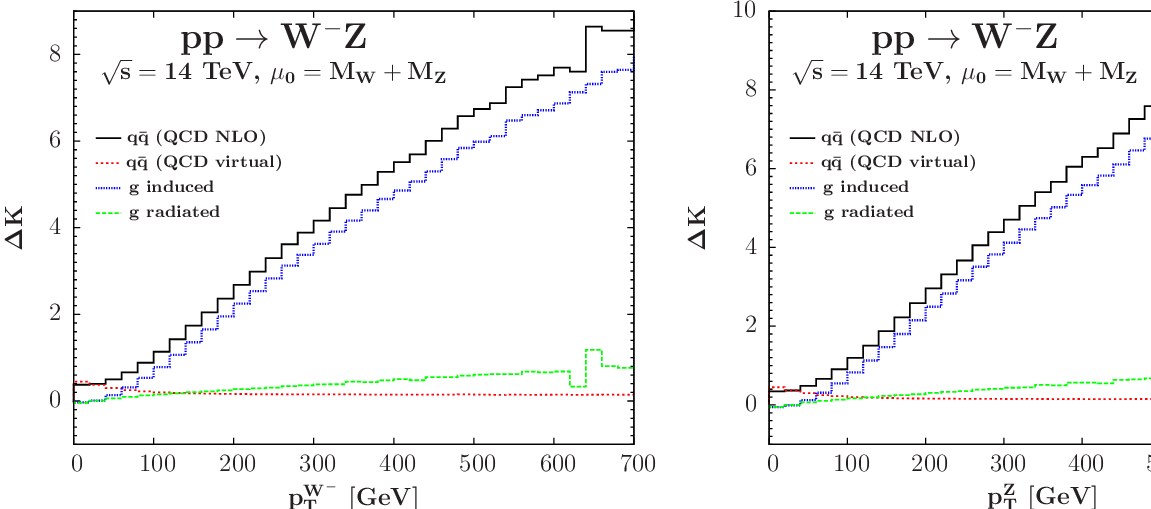}
\end{center}
\it{\vspace{-6mm}\caption{\small $W$ and $Z$ transverse momentum $p_T$
    (in TeV) distributions of the NLO QCD corrections (using $\Delta K$) to $pp\to W^+Z$
    (upper left and right respectively) and $pp\to W^-Z$ (lower left
    and right respectively) cross sections at the LHC,
    calculated with {\texttt{MRST2004QED}} PDF set and with the input parameters
    described in
    Section~\protect\ref{section-parameters}.\label{WZ-QCD-pT-distribution}}}
\end{figure}

In the case of the NLO QCD corrections we observe the expected behavior
with similar shape for both distributions in $W^+Z$ and $W^-Z$
channels, as displayed in \fig{WZ-QCD-pT-distribution}. The QCD
corrections are up to $\Delta K \sim 7$ at $p_T^{W^+}\sim 700$ GeV and
$p_T^{Z}\sim 700$ GeV in the $W^+Z$ channel and larger in the $W^-Z$
channel with corrections up to $\Delta K \sim 8$ at $p_T^{W^-}\sim
700$ GeV and $\Delta K \sim 9$ at $p_T^Z\sim 700$ GeV. Similar to the
$ZZ$ case, those huge corrections can be explained as follows. For the
$p_T^{W^+}$ distribution the dominant mechanism is $ug\to W^+ d$ and
then a soft $Z$ is radiated from a quark or the $W^+$.  For $p_T^{W^+}
\gg M_Z$ we have, with $a_W = 1/(\sqrt{2}\sin\theta_W)$,
\begin{align}
d\sigma^{ug \to W^+Z d}
&= \frac12 c_{L,d}c_{L,u}\left(1 + \fr{\cot\theta_W}{c_{L,d}} -
  \fr{\cot\theta_W}{c_{L,u}}\right) d\sigma_L^{ug \to W^+ d}
\fr{\alpha}{2\pi}\log^2\left[\fr{(p_T^{W^+})^2}{M_Z^2}\right]\crn
&= c^u_{WZ} d\sigma_L^{ug \to Z u}
\fr{\alpha}{2\pi}\log^2\left[\fr{(p_T^{W^+})^2}{M_Z^2}\right], \crn
c^u_{WZ} &= \frac12 a_W^2 \fr{c_{L,d}}{c_{L,u}}\left(1 + \fr{\cot\theta_W}{c_{L,d}}
  - \fr{\cot\theta_W}{c_{L,u}}\right) = 4.13,
\label{eq_log2_wpz}
\end{align}
where the $\bar{d}g$ contribution, about an order of magnitude smaller
than the $ug$ one, has been neglected. We also have used the following
relations for amplitudes at high energies
\begin{align}
\mathcal{A}_L^{ug \to W^+ d} &= \fr{a_W}{c_{L,u}}\mathcal{A}_L^{ug \to
  Z u},\crn
\mathcal{A}_L^{dg \to W^- u} &= \fr{a_W}{c_{L,d}}\mathcal{A}_L^{dg \to
  Z d},\crn
\mathcal{A}_L^{dg \to Z d} &= \fr{c_{L,d}}{c_{L,u}}\mathcal{A}_L^{ug
  \to Z u}.
\label{eq_qg_wz}
\end{align}
Doing the same exercise for the $p_T^{W^-}$
distribution we get
\begin{align}
d\sigma^{dg \to W^-Z u}
&= \frac12 c_{L,d}c_{L,u}\left(1 + \fr{\cot\theta_W}{c_{L,d}} -
  \fr{\cot\theta_W}{c_{L,u}}\right) d\sigma_L^{dg \to W^-u}
\fr{\alpha}{2\pi}\log^2\left[\fr{(p_T^{W^-})^2}{M_Z^2}\right]\crn
&= c^d_{WZ} d\sigma_L^{dg \to Z d}
\fr{\alpha}{2\pi}\log^2\left[\fr{(p_T^{W^-})^2}{M_Z^2}\right],\crn
c^d_{WZ} &= \frac12 a_W^2 \fr{c_{L,u}}{c_{L,d}} \left(1 +
  \fr{\cot\theta_W}{c_{L,d}} - \fr{\cot\theta_W}{c_{L,u}}\right) = 2.81.
\label{eq_log2_wmz}
\end{align}
The huge correction for the $p_T^Z$ distributions can be explained
using the same arguments, but we have to consider soft $W^\pm$
radiation instead of soft $Z$ radiation. One has to be careful with
the coupling constants because radiating a $W^\pm$ changes the quark
flavor. For $p_T^Z >> M_W$ we get
\begin{align}
d\sigma^{ug \to W^+Z d}
&= c^u_{WZ} d\sigma_L^{ug \to Z u}
\fr{\alpha}{2\pi}\log^2\left[\fr{(p_T^{Z})^2}{M_W^2}\right],\crn
d\sigma^{dg \to W^-Z u}
&= c^d_{WZ} d\sigma_L^{dg \to Z d}
\fr{\alpha}{2\pi}\log^2\left[\fr{(p_T^{Z})^2}{M_W^2}\right],
\label{eq_log2_wpmz_pTZ}
\end{align}
for $W^+ Z$ and $W^- Z$ production, respectively. This result can be
obtained from Eqs.~(\ref{eq_log2_wpz},\ref{eq_log2_wmz}) by replacing
$M_Z$ with $M_W$. This explains why the corrections for the $p_T^Z$
distributions are a little bit larger than the corresponding
$p_T^{W^\pm}$ ones. We observe that \eq{eq_log2_wpmz_pTZ} differs from
the result of Ref.~\cite{WZ-QCD-NLO-2}.
Numerically, we get $\Delta K_{qg} = 12.61$ ($6$) and $17.22$ ($7.60$)
for leading-logarithmic approximation (full calculation)
for $p_T^{W^+}$ and $p_T^{W^-}$ distributions, respectively, at $p_T=700$ GeV.
For $p_T^Z$ distributions, we get $\Delta K_{qg} = 14.22$ ($6.30$) and
$19.42$ ($9.00$) for leading-logarithmic approximation (full
calculation) for $W^+Z$ and $W^-Z$, respectively.

\begin{figure}[ht!]
\begin{center}
\includegraphics[scale=0.7]{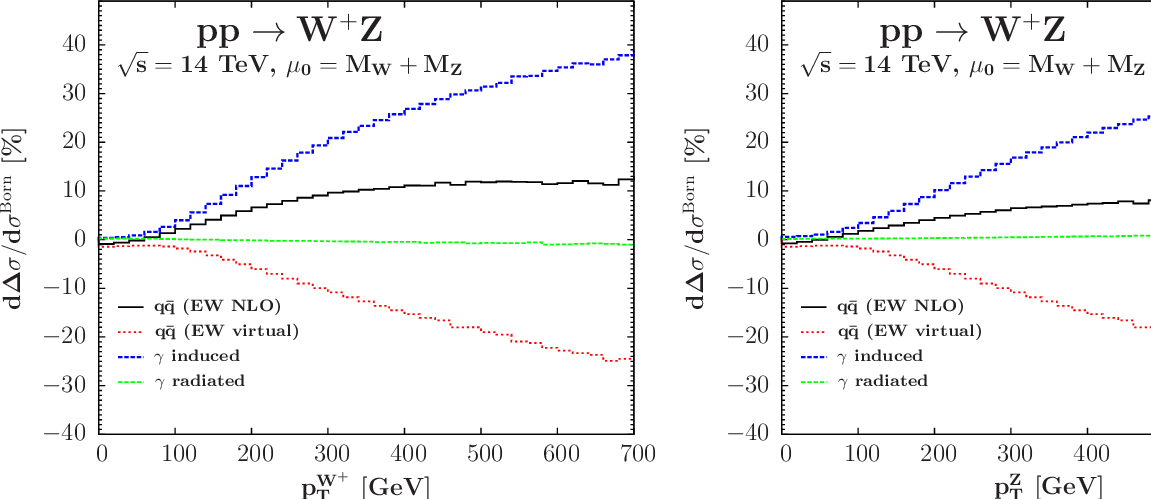}

\vspace{2mm}
\includegraphics[scale=0.7]{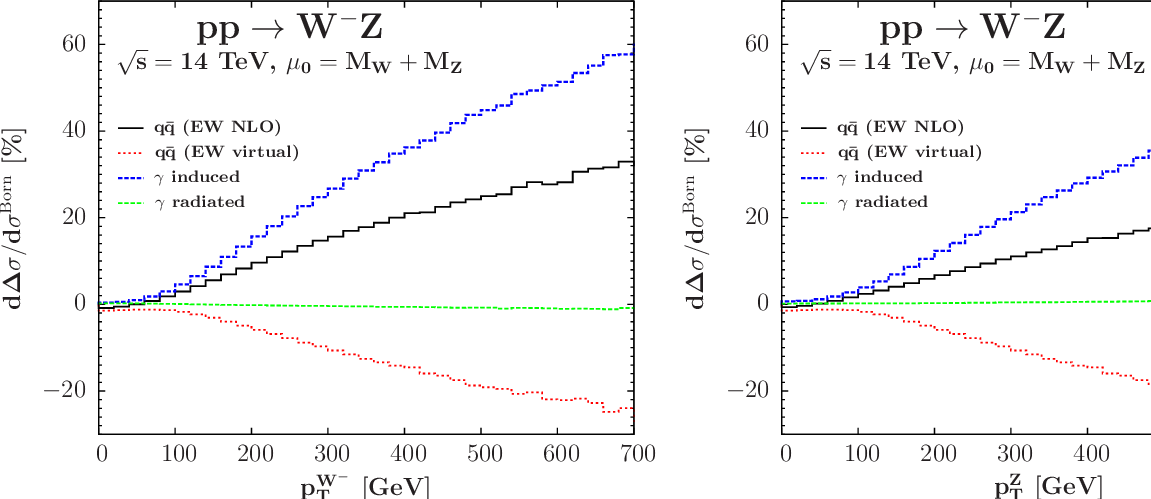}
\end{center}
\it{\vspace{-6mm}\caption{\small Same as
    \protect\fig{WZ-QCD-pT-distribution} but for EW
    corrections (in $\%$).\label{WZ-EW-pT-distribution}}}
\end{figure}

As for the EW effects, the NLO corrections for the transverse momentum
distributions are shown in \fig{WZ-EW-pT-distribution}.  In both
channels and for both transverse momenta the effect of the photon
radiated processes is negligible. Comparing to the $ZZ$ case, we
observe that the virtual correction is significantly less
negative. This suggests that there are more cancellations between
negative double-logarithm and positive single-logarithm corrections in
the $W^\pm Z$ cases. More striking is the difference in the
photon-quark induced corrections, it is $+60\%$ for the $p_T^{W^-}$
distribution, while only $+0.3\%$ for the $ZZ$ case, at
$700~\text{GeV}$.
The difference between $W^+Z$ and $W^-Z$ channels is also clearly
visible: in the upper panels the photon-quark induced processes (in dotted
dashed blue) in the $W^+Z$ channel compensate nearly exactly the
effect of the virtual corrections (in dotted red) reducing the total
EW corrections to less than $\sim +10\%$ on the whole transverse
momentum range both for the $W$ and $Z$ bosons, while in the case of
$W^-Z$ process the photon-quark induced corrections are larger, driving the
EW corrections to $\sim +30\%$ for $p_T^{W^-}\sim 700$ GeV and $\sim
+20\%$ for $p_T^{Z}\sim 700$ GeV. In both channels the difference with
the $ZZ$ channel is much more enhanced than in the QCD case, and
the key difference is that the $W^\pm$ can couple to the
photon in the EW case. This introduces a new Feynman diagram with
$t$-channel $W^\pm$ exchange in the $2\to 2$ process and a new
possibility of radiating a soft $W^\pm$ from the initial-state photon. A
detailed calculation is presented in the Appendix with all the
intermediate steps. To explain the $p_T^{Z}$ distribution we have to
consider soft $W^\pm$ radiation as in the QCD case. For $p_T^{Z} \gg
M_W$ we get
{\small \begin{align}
d\sigma^{u\gamma \to W^+Z d}
&= \left[\frac12 a_W^2(1-a_u+\fr{a_u^2}{2})d\sigma_L^{u\gamma \to Z u}
+ \fr{1}{4}\cot^2\theta_W d\sigma_{L}^{u\gamma \to W^+ d}
+ \fr{1}{4}d\sigma_{LT}^{uW_{\gamma}^- \to Z d} \right]
\fr{\alpha}{2\pi}\log^2\left[\fr{(p_T^{Z})^2}{M_W^2}\right],\crn
d\sigma^{d\gamma \to W^-Z u}
&= \left[\frac12 a_W^2(1-a_d+\fr{a_d^2}{2})\, d\sigma_L^{d\gamma \to Z d}
+ \fr{1}{4}\cot^2\theta_W d\sigma_{L}^{d\gamma \to W^- u}
+ \fr{1}{4}d\sigma_{LT}^{dW_{\gamma}^+ \to Z u} \right]
\fr{\alpha}{2\pi}\log^2\left[\fr{(p_T^{Z})^2}{M_W^2}\right],
\label{eq_log2_wz_EW_softW}
\end{align}}
where $W_{\gamma}^\pm$ means that the photon PDF must be used and
\begin{align}
a_u = 1 - \fr{Q_dc_{L,d}}{Q_uc_{L,u}},\;\; a_d = 1 -
\fr{Q_uc_{L,u}}{Q_dc_{L,d}}.
\label{eq:au_ad}
\end{align}
In order to obtain the results in \eq{eq_log2_wz_EW_softW} we have
used the following identities, which are true in the high-energy
limit,
\begin{align}
\cot\theta_W\mathcal{A}_{L}^{u\gamma \to W^+ d} - \mathcal{A}_{LT}^{uW^-
  \to Z d} &= a_Wa_u \mathcal{A}_L^{u\gamma \to Z u},\crn
\cot\theta_W\mathcal{A}_{L}^{d\gamma \to W^- u} - \mathcal{A}_{LT}^{dW^+
  \to Z u} &= -a_Wa_d \mathcal{A}_L^{d\gamma \to Z d},
\label{eq_amp_wz}
\end{align}
where all the gauge bosons are transverse. This is because the
longitudinal-mode contributions to \eq{eq_log2_wz_EW_softW} vanish in
the high-energy limit $p_T \gg M_Z$. Therefore, all the gauge bosons are
transverse in all leading-logarithmic results presented in this
paper. More details are given in the Appendix.
For $p_T^{W^\pm} \gg M_Z$ with soft $Z$ radiation we have
\begin{align}
d\sigma^{u\gamma \to W^+Z d}
&= \fr{c_{L,u}^2 c^u_{WZ}}{a_W^2}d\sigma_L^{u\gamma \to W^+
    d}
\fr{\alpha}{2\pi}\log^2\left[\fr{(p_T^{W^+})^2}{M_Z^2}\right],\crn
d\sigma^{d\gamma \to W^-Z u}
&= \fr{c_{L,d}^2 c^d_{WZ}}{a_W^2}d\sigma_L^{d\gamma \to W^-
    u}
\fr{\alpha}{2\pi}\log^2\left[\fr{(p_T^{W^-})^2}{M_Z^2}\right].
\label{eq_log2_wz_EW_softZ}
\end{align}

The main reason why the photon-quark induced corrections are much larger in
the $WZ$ case than in the $ZZ$ case is because the cross sections
$d\sigma_L^{d\gamma \to W^-u}$, $d\sigma_L^{u\gamma \to W^+d}$,
$d\sigma_{LT}^{dW_{\gamma}^+ \to Zu}$ and $d\sigma_{LT}^{uW_{\gamma}^- \to Zd}$,
involving a $t$-channel Feynman diagram with a gauge boson exchange,
are about one to two orders of magnitude larger than $d\sigma_L^{u\gamma \to Zu}$ at
$p_T^Z\approx 700$ GeV. Qualitatively, this can be
  understood as follows. Considering the ratio
of the $t$-channel Feynman diagram with a gauge boson exchange in 
$u\gamma \to W^+d$
to the $t$-channel Feynman diagram with a quark exchange in $u\gamma \to 
Z u$, we get
the factor $E_{\gamma}/|q|$ with $q^2 \approx -2E^2_{\gamma}(1- 
\cos\theta)$ from dimensional analysis.
At the amplitude squared level, the factor becomes $1/[2(1- 
\cos\theta)]$, which
is about $8$ for $p_T = 700$~GeV and $E_{\gamma} = 2$~TeV. Here we are 
assuming that
the dominant contribution comes from the region of $4$~TeV of partonic 
center of mass energy.
This is reasonable because, compared to $u\gamma \to Zu$, the
Feynman parameters $x_i$ ($i=1,2$)
(for dominant contribution) in $u\gamma \to W^+d$ are expected to be 
larger due to the exchange of
a $t$-channel gauge boson. Some further enhancement
from the couplings can be possible as we have seen in the previous QCD 
results. Numerically, we get $37.3\%$ ($38\%$), $69.5\%$ ($58\%$)
for leading-logarithmic approximation (full calculation)
for $p_T^{W^+}$ and $p_T^{W^-}$ distributions, respectively, at $p_T=700$ GeV.
For $p_T^Z$ distributions, we get
$58.8\%$ ($32\%$) and $100\%$ ($48\%$)
for leading-logarithmic approximation (full calculation) for $W^+Z$
and $W^-Z$, respectively.

\begin{figure}[ht!]
\begin{center}
\hspace{-1.2cm}\includegraphics[scale=0.7]{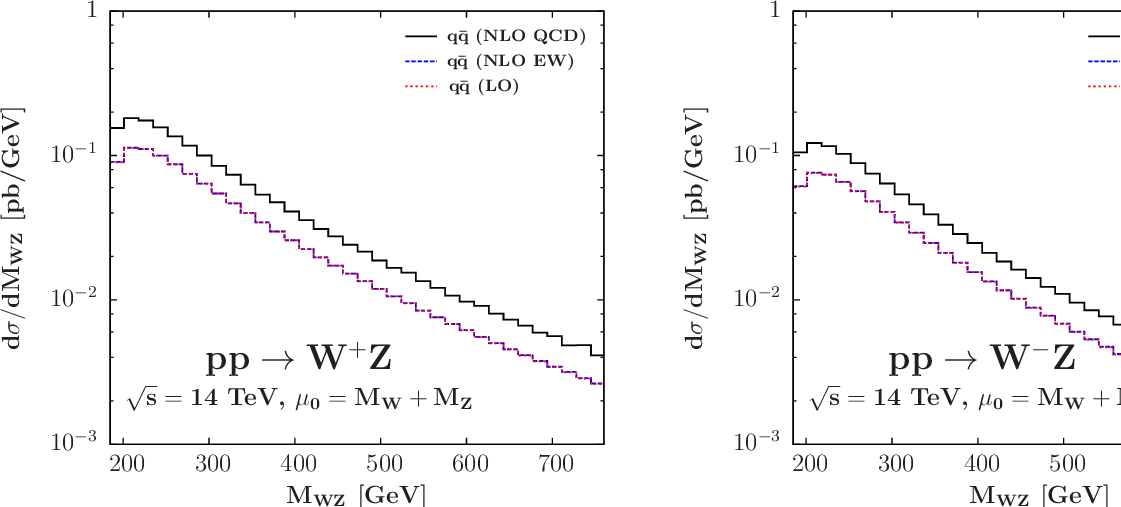}

\vspace{2mm}
\includegraphics[scale=0.7]{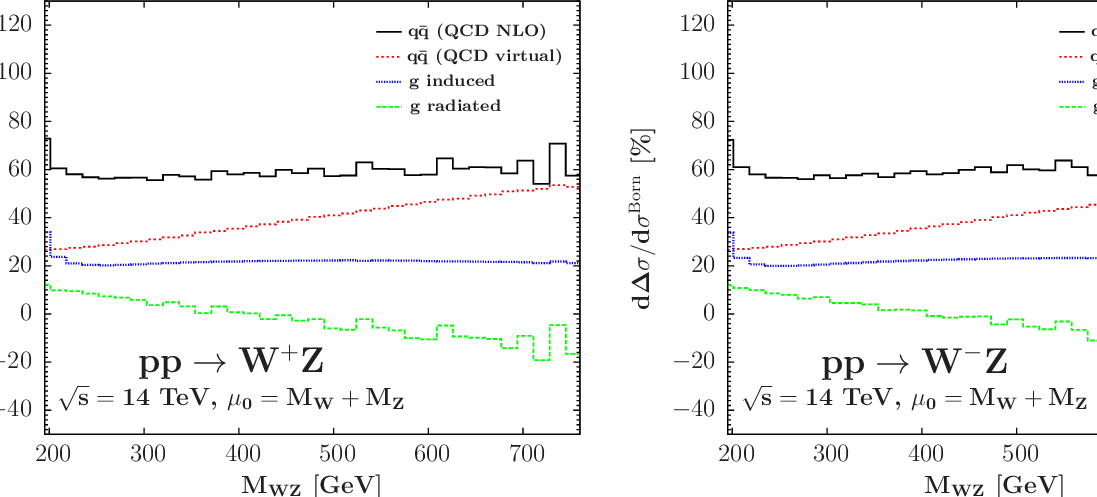}

\vspace{2mm}
\includegraphics[scale=0.7]{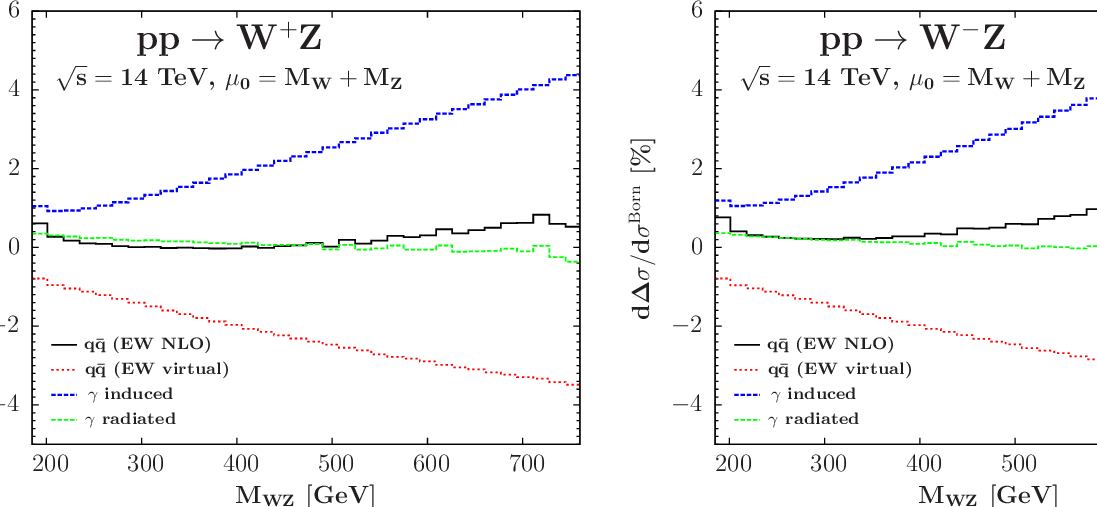}
\end{center}
\it{\vspace{-6mm}\caption{\small $M_{W\!Z}$ invariant mass (in GeV)
    distribution of $pp\to W^+Z$ (upper left) and $pp\to W^-Z$ (upper
    right) cross sections at the LHC (in pb/GeV), including NLO QCD and EW
    corrections calculated with {\texttt{MRST2004QED}} PDF set and with the input
    parameters described in
    Section~\protect\ref{section-parameters}. The $q\bar{q}$ LO
    (dotted red) and $q\bar{q}$ NLO EW (dashed blue) curves nearly
    coincide. Middle left (right) panels: $M_{W\!Z}$ invariant mass
    distribution of the NLO QCD corrections to $pp\to W^+Z$ ($pp\to
    W^-Z$) cross section at the LHC (in $\%$) ; lower left and right:
    the same but for the NLO EW
    corrections.\label{WZ-mWZ-distribution}}}
\end{figure}

The invariant mass distributions are similar in $W^+Z$ and $W^-Z$
channels. They are displayed in \fig{WZ-mWZ-distribution}. The QCD
corrections, displayed in middle panels, are of the order of
$+60\%$ (with some fluctuations up to $+70\%$ near the threshold). As
displayed in lower panels of \fig{WZ-mWZ-distribution} the EW
corrections are very small, $\sim +1\%$ in $W^+Z$ channel for
$M_{W\!Z}\geq 500$ GeV (and close to zero just after the threshold) and
$\sim +2\%$ in $W^-Z$ at $M_{W\!Z}\sim 700$ GeV. We again see a
compensation of the virtual Sudakov logarithm (in dotted red), giving
a decreasing virtual corrections down to $-4\%$ in both channels, to
be added to the positive correction coming from the photon-quark induced
processes (dotted dashed blue) up to $\sim +4\%$ in $W^+Z$ channel and
$\sim +5\%$ in $W^-Z$ channel.

\begin{figure}[ht!]
\begin{center}
\includegraphics[scale=0.72]{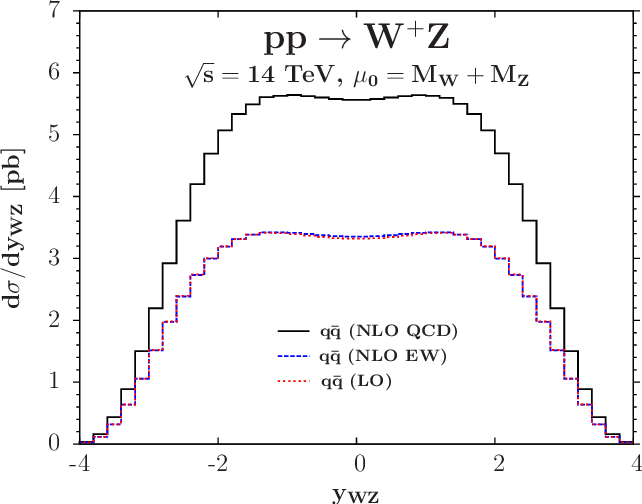}
\hspace{6mm}
\includegraphics[scale=0.72]{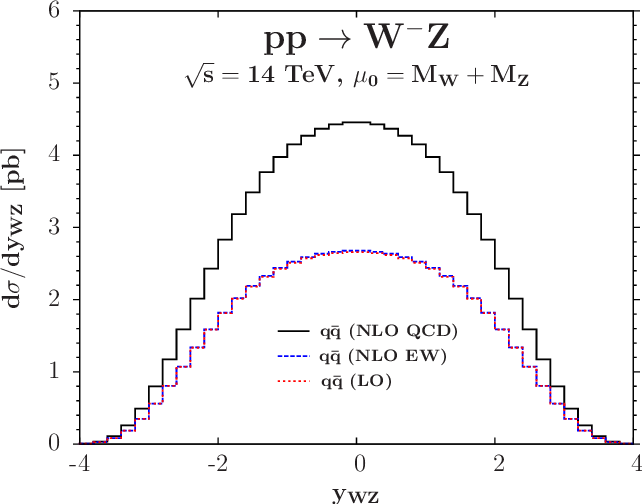}

\vspace{2mm}
\includegraphics[scale=0.72]{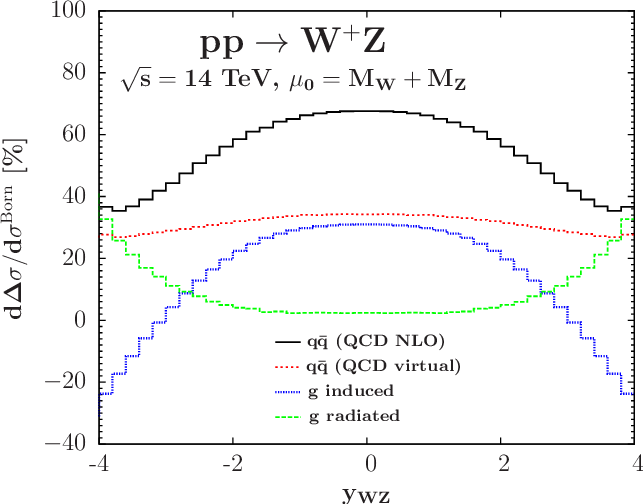}
\hspace{6mm}
\includegraphics[scale=0.72]{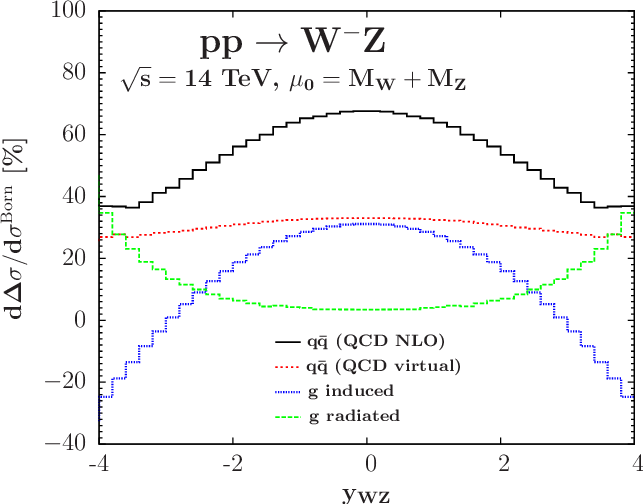}

\vspace{2mm}
\includegraphics[scale=0.72]{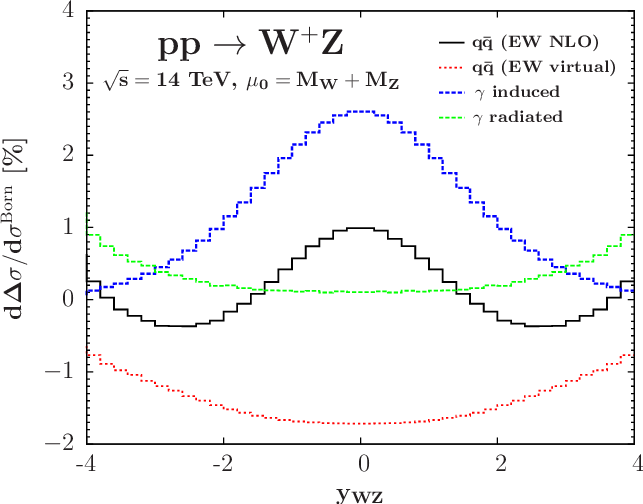}
\hspace{6mm}
\includegraphics[scale=0.72]{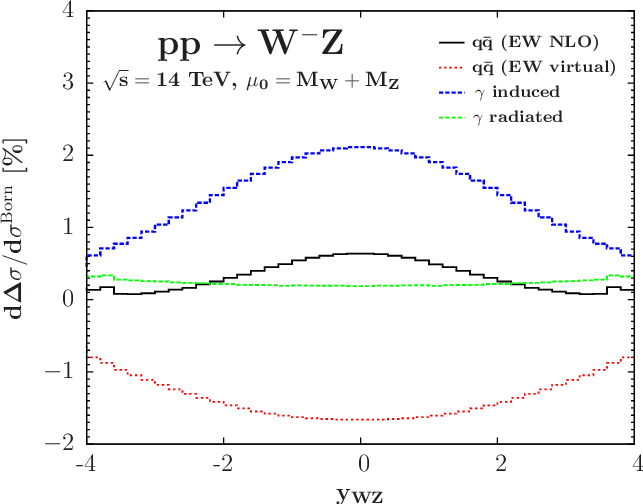}
\end{center}
\it{\vspace{-6mm}\caption{\small $WZ$ pair rapidity $y_{W\!Z}$
    distributions of $pp\to W^+Z$ (upper left, in pb) and $pp\to W^-Z$
    (upper right, in pb) as well as the $y_{W\!Z}$ distributions of the
    NLO QCD corrections and the NLO EW corrections in $pp\to W^+Z$
    channel (middle left and lower left, respectively, in $\%$) and in
    $pp\to W^-Z$ channel (middle right and lower right, respectively),
    at the LHC, calculated with {\texttt{MRST2004QED}} PDF set and with the input
    parameters described in
    Section~\protect\ref{section-parameters}. The $q\bar{q}$ LO
    (dotted red) and $q\bar{q}$ NLO EW (dashed blue) curves nearly
    coincide in the upper panels.\label{WZ-rapidity-distributions}}}
\end{figure}

We end the study of $W^+Z$ and $W^-Z$ channels with the rapidity
distributions of the $WZ$ pair displayed in
\fig{WZ-rapidity-distributions}. $W^+Z$ and $W^-Z$ channels have
a slightly different shape, the former channel having a double
gaussian shape with a minimum around $y_{WZ}=0$ while the latter displays a
gaussian shape with a maximum around $y_{WZ}=0$. The EW corrections are
negligible, no more than $\sim +3\%$ ($\sim +2\%$) in $W^+Z$ channel
($W^-Z$ channel), as depicted in the lower panels of
\fig{WZ-rapidity-distributions}, compared to the $\sim
+40$--$60\%$ increase due to the QCD corrections as seen in the
middle panels. The photon-quark induced processes play a very important
role in the EW corrections: while the virtual corrections are always
negative from $\sim -1\%$ down to $\sim -2\%$ at $y_{WZ}=0$ in both
channels, the photon-quark induced processes are always positive with a
gaussian shape in both channels, resulting in a mexican hat shape for
the EW corrections, sharper in the $W^+Z$ channel than in the $W^-Z$
channel.

\subsection{\texorpdfstring{$WW$}{WW}
  distributions\label{section-WW}}

\begin{figure}[ht!]
\begin{center}
\includegraphics[scale=0.7]{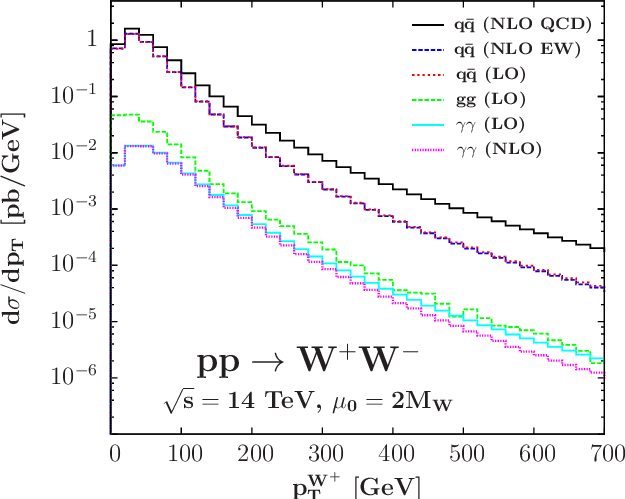}
\hspace{10mm}
\includegraphics[scale=0.7]{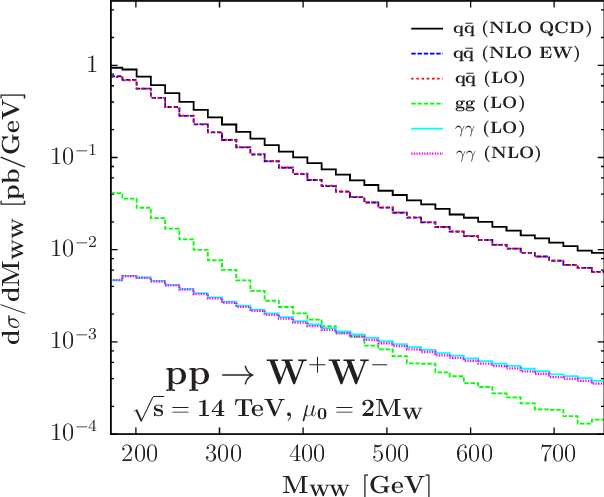}
\end{center}
\it{\vspace{-6mm}\caption{\small $W^+$ transverse momentum $p_T$ (in GeV)
    distribution (left) and $W$ pair invariant mass $M_{W\!W}$ (in GeV)
    distribution (right) of $pp\to WW$ cross section at the LHC (in
    pb/GeV), including NLO QCD and EW corrections calculated with
    {\texttt{MRST2004QED}} PDF set and with the input parameters described in
    Section~\protect\ref{section-parameters}. The LO $q\bar{q}$ (dotted red) and
    NLO EW $q\bar{q}$ (dashed blue) curves nearly coincide, as well
    as the LO $\gamma\gamma$ (solid light blue) and NLO
    $\gamma\gamma$ (dotted violet) in the $M_{WW}$
    distribution.\label{WW-all-distribution}}}
\end{figure}

\begin{figure}[ht!]
\begin{center}
\includegraphics[scale=0.7]{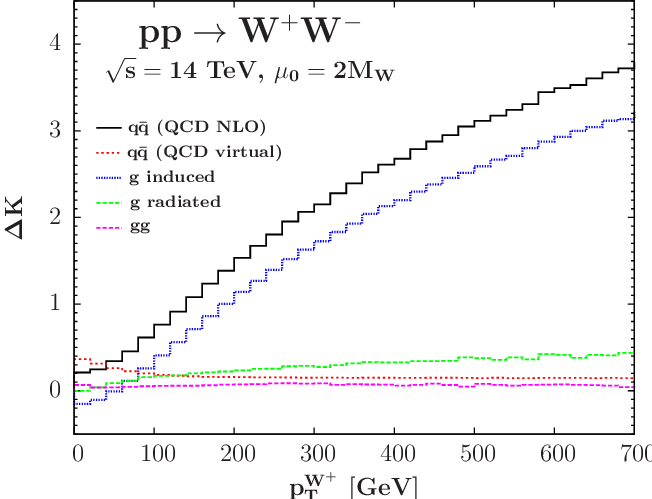}
\hspace{10mm}
\includegraphics[scale=0.7]{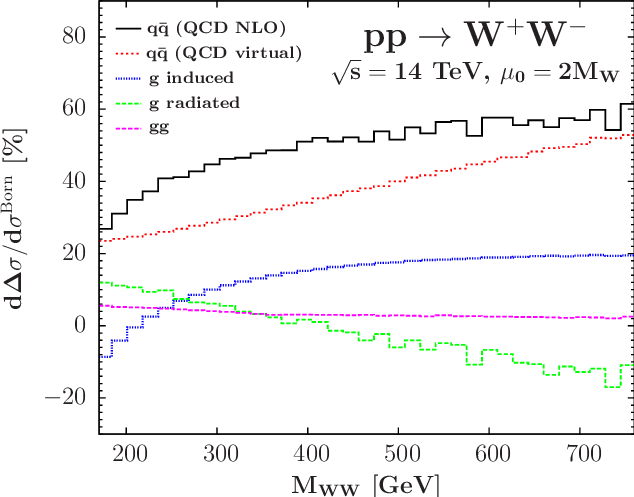}
\end{center}
\it{\vspace{-6mm}\caption{\small $W^+$ transverse momentum $p_T$ (in TeV)
    distribution (left, using $\Delta K$) and $W$ pair invariant mass $M_{W\!W}$ (in GeV)
    distribution (right, in $\%$) of the NLO QCD corrections to $pp\to WW$
    cross section at the LHC, calculated with {\texttt{MRST2004QED}}
    PDF set and with the input parameters described in
    Section~\protect\ref{section-parameters}. \label{WW-QCD-distribution}}}
\end{figure}

We do the same exercise as above for $pp\to W^+W^-$ channel. The
distributions of the transverse momentum $p_T^{W^+}$ of the $W^+$ and
the $WW$ invariant mass are shown in \fig{WW-all-distribution} and
show a similar behavior as for $ZZ$ production. Notable is the
contribution of the $\gamma\gamma$ induced process. In the $p_T^{W^+}$
distribution it is only slightly smaller than the $gg$ contribution
except at small $p_T$. Its contribution to the $WW$ invariant mass
distribution is about one order of magnitude smaller than the $gg$
induced process at low invariant masses while it is about a factor 2
larger at $M_{WW} \simeq 700$ GeV. The trend that is seen in
\fig{WW-QCD-distribution} for the $p_T^{W^+}$ and $M_{W\!W}$
distributions of the QCD corrections is the same as the one seen in
\fig{ZZ-QCD-distribution} for $ZZ$ production. The QCD corrections in
the $p_T^{W^+}$ distribution are driven by a double-logarithmic
enhancement in the gluon-quark induced subprocesses and one has up to a
$\Delta K \sim 3$ correction at $p_T^{W^+}=700~\text{GeV}$.

Similar to the $ZZ$ and $WZ$ cases, to explain the $p_T^{W^+}$
distribution we have to consider soft $W^-$ radiation. For
$p_T^{W^+}\gg M_W$, keeping only the leading corrections, we get
\begin{align}
d\sigma^{qg \to W^+W^- q}
&= c^q_{WW}d\sigma_L^{qg \to Z q}
\fr{\alpha}{2\pi}\log^2\left[\fr{(p_T^{W^+})^2}{M_W^2}\right],\crn
c^q_{WW} &= \fr{a_W^4}{2 c^2_{L,q}},\;\;\; q = u, d.
\label{eq_log2_ww_pTWpm}
\end{align}
For the sake of comparison with $c^q_{ZZ}$ and $c^q_{WZ}$, we have
$c_{WW}^u = 3.53$ and $c_{WW}^d = 2.40$. Numerically, for the
$p_T^{W^+}$ distribution, we get $\Delta K_{qg} = 5.33$ ($3.10$) for
leading-logarithmic approximation (full calculation) at $p_T=700$
GeV.

In the invariant mass distribution the gluon-quark induced and the
gluon-quark radiated processes compensate each other and the full QCD
corrections are dominated by the virtual corrections. The gluon fusion
mechanism amounts to $\sim +5\%$ in the $M_{W\!W}$ distribution and is
negligible in the transverse momentum distribution.

\begin{figure}[ht!]
\begin{center}
\includegraphics[scale=0.7]{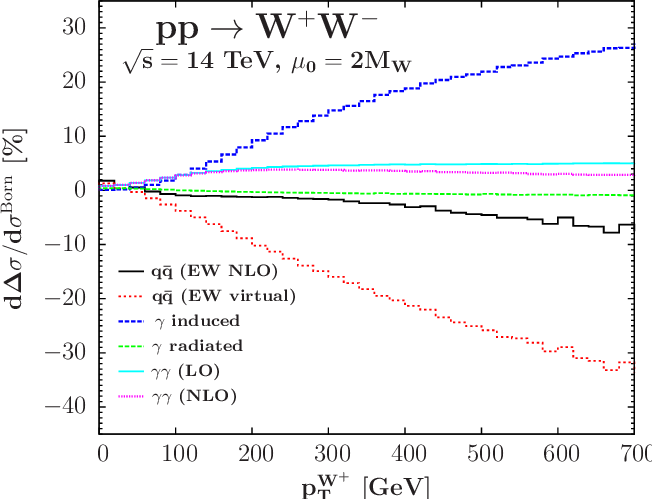}
\hspace{10mm}
\includegraphics[scale=0.7]{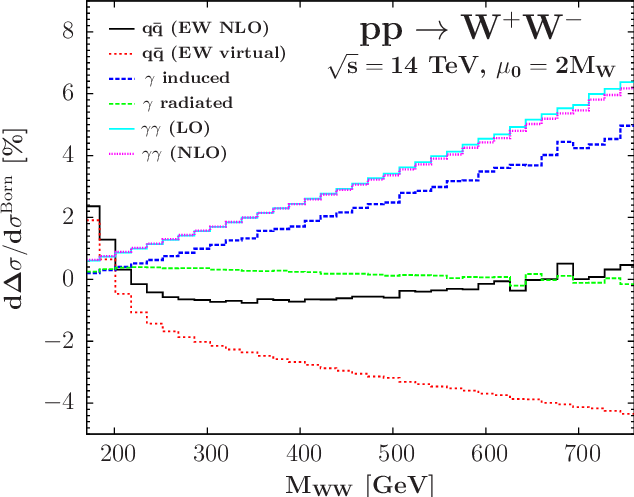}
\end{center}
\it{\vspace{-6mm}\caption{\small Same as
    \protect\fig{WW-QCD-distribution} but for EW
    corrections (in $\%$). The LO $\gamma\gamma$ (solid light blue) and
    NLO $\gamma\gamma$ (dotted violet) curves coincide in the
    $M_{WW}$ distribution.\label{WW-EW-distribution}}}
\end{figure}

Turning to EW corrections, we observe that the effect of the virtual
Sudakov logarithms $\alpha \log^2[(p_T^{W^+})^2/M_W^2]$ is clearly
visible in \fig{WW-EW-distribution} (left) where the dotted red curve
is the $p_T^{W^+}$ distribution of the virtual EW corrections, and we
have a $-30\%$ correction at $p_T^{W^+}=600$ GeV. The
interesting point is again that this Sudakov logarithm is compensated
by the effect of photon-quark induced processes displayed in dotted-dashed
blue as already seen in the case of $WZ$
distributions. Eventually the EW corrections in the $q\bar q$
subprocesses are limited and not more than $-10\%$ over the whole $p_T$
range. The same effect is visible in the invariant mass distribution,
where the total EW corrections to the $q\bar q$ subprocesses drop very
fast from $+2\%$ at the threshold to small variations around zero. As
in the case of $WZ$, by considering the limit of soft $W^-$
radiation we get for $p_T^{W^+}\gg M_W$
\begin{align}
d\sigma^{u\gamma \to W^+W^- u} &=
\left(\fr{a_W^4}{4c^2_{L,u}}d\sigma_L^{u\gamma \to Z u} +
  \fr{a_W^2}{4}d\sigma_L^{u\gamma \to W^+ d} +
  \fr{1}{4}d\sigma_{LT}^{uW_{\gamma}^+ \to W^+ u}\right)
\fr{\alpha}{2\pi} \log^2\left[\fr{(p_T^{W^+})^2}{M_W^2}\right],\crn
d\sigma^{d\gamma \to W^+W^- d} &=
\left(\fr{a_W^4}{4c^2_{L,d}}d\sigma_L^{d\gamma \to Z d} +
  \fr{a_W^2}{4}d\sigma_L^{u_{d}\gamma \to W^+ d} +
  \fr{1}{4}d\sigma_{LT}^{dW_{\gamma}^+ \to W^+ d}\right)
\fr{\alpha}{2\pi} \log^2\left[\fr{(p_T^{W^+})^2}{M_W^2}\right],
\label{eq_log2_ww_softWm}
\end{align}
where $u_{d}$ means that the $d$ PDF must be used. And for the limit of
soft $W^+$ radiation with $p_T^{W^-}\gg M_W$ we have
\begin{align}
d\sigma^{u\gamma \to W^+W^- u} &=
\left(\fr{a_W^4}{4c^2_{L,u}}d\sigma_L^{u\gamma \to Z u} +
  \fr{a_W^2}{4}d\sigma_L^{d_{u}\gamma \to W^- u} +
  \fr{1}{4}d\sigma_{LT}^{uW_{\gamma}^- \to W^- u}\right)
\fr{\alpha}{2\pi} \log^2\left[\fr{(p_T^{W^-})^2}{M_W^2}\right],\crn
d\sigma^{d\gamma \to W^+W^- d} &=
\left(\fr{a_W^4}{4c^2_{L,d}}d\sigma_L^{d\gamma \to Z d} +
  \fr{a_W^2}{4}d\sigma_L^{d\gamma \to W^- u} +
  \,\,\,\, \fr{1}{4}d\sigma_{LT}^{dW_{\gamma}^- \to W^- d}\right)
\fr{\alpha}{2\pi} \log^2\left[\fr{(p_T^{W^-})^2}{M_W^2}\right],
\label{eq_log2_ww_softWp}
\end{align}
where $d_{u}$ means that the $u$ PDF must be used. Similar to the
$WZ$ case, only the transverse gauge bosons contribute and we
have used the following identities in the high-energy limit,
\begin{align}
a_W\mathcal{A}_L^{u\gamma \to W^+ d} - \mathcal{A}_{LT}^{uW^+ \to W^+ u}
&= \fr{a_W^2}{c_{L,u}} \mathcal{A}_L^{u\gamma \to Z u},\crn
a_W\mathcal{A}_L^{d\gamma \to W^- u} - \mathcal{A}_{LT}^{uW^- \to W^- u}
&= \fr{a_W^2}{c_{L,u}} \mathcal{A}_L^{u\gamma \to Z u},\crn
a_W\mathcal{A}_L^{d\gamma \to W^- u} + \mathcal{A}_{LT}^{dW^- \to W^- d}
&= \fr{a_W^2}{c_{L,d}} \mathcal{A}_L^{d\gamma \to Z d},\crn
a_W\mathcal{A}_L^{u\gamma \to W^+ d} + \mathcal{A}_{LT}^{dW^+ \to W^+ d}
&= \fr{a_W^2}{c_{L,d}} \mathcal{A}_L^{d\gamma \to Z d}.
\label{eq_amp_ww}
\end{align}
Numerically, for the $p_T^{W^+}$ distribution, we get $22\%$ ($26\%$)
for leading-logarithmic approximation (full calculation) at
$p_T=700$ GeV. As in the $WZ$ case, the large photon-quark induced
corrections are due to the hard $2\to 2$ amplitudes with a $t$-channel
gauge boson exchange.

The $\gamma\gamma$ subprocess has more impact on the invariant mass
distribution than on the transverse momentum one, increasing up to
$+6\%$ at about $700$ GeV. As expected, the NLO corrections in the
$\gamma\gamma$ subprocess are dominated by the negative
Sudakov-logarithm corrections from the virtual part, as can be seen in
the $p_T^{W^+}$ distribution.

\begin{figure}[ht!]
\begin{center}
\includegraphics[scale=0.7]{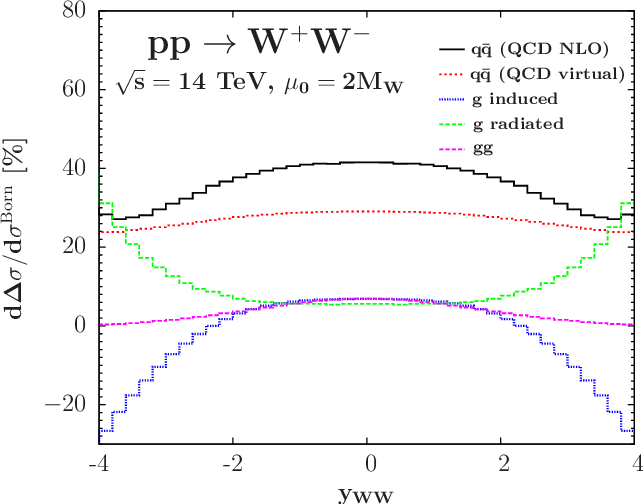}
\hspace{6mm}
\includegraphics[scale=0.7]{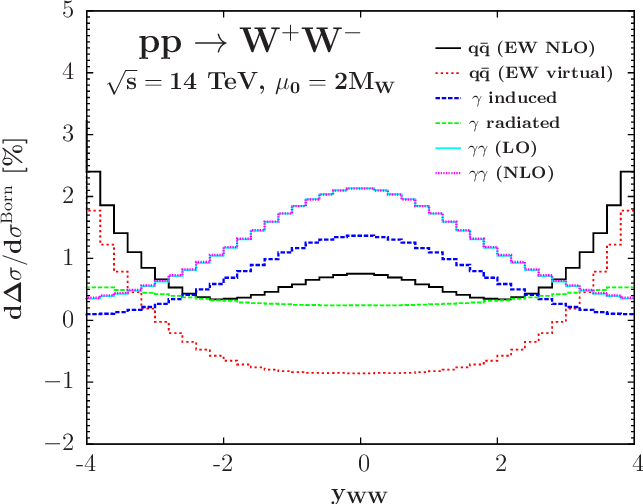}

\vspace{3mm}
\includegraphics[scale=0.7]{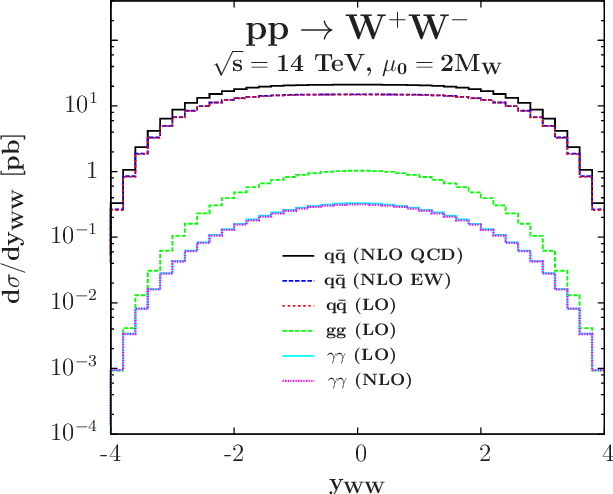}
\end{center}
\it{\vspace{-6mm}\caption{\small $W$ pair rapidity $y_{W\!W}$
    distribution of the NLO QCD corrections and NLO EW corrections
    (upper left and right, respectively, in $\%$), as well as the
    distribution of the $pp\to WW$ cross section (lower, in pb) at
    the LHC, calculated with {\texttt{MRST2004QED}} PDF set and with the input
    parameters described in
    Section~\protect\ref{section-parameters}. The LO $\gamma\gamma$
    (solid light blue) and NLO $\gamma\gamma$ (dotted violet) curves
    coincide as well as the LO $q\bar{q}$ (dotted red) and NLO EW
    $q\bar{q}$ (dashed blue) in the lower
    panel.\label{WW-rapidity-distributions}}}
\end{figure}
We also display the rapidity distribution of the $W$ pair in
\fig{WW-rapidity-distributions}. As can be
seen on the lower panel of \fig{WW-rapidity-distributions} the
impact of NLO corrections on the diphoton sub-channel is negligible in the
rapidity distribution. The impact of the diphoton and gluon fusion
channels are two to three order of magnitudes less than the NLO QCD
corrections to the $q\bar{q}$ sub-channels, as already seen in previous
analyses, while the total EW corrections (including diphoton
corrections) are limited as they amount to less than $\sim +2\%$ and
are always positive. The upper right panel of
\fig{WW-rapidity-distributions} gives another example of the
important impact of the photon-quark induced processes in the structure of
the EW corrections on $W$ pair production: while the EW virtual
corrections display a negative quadratic behavior with a minimum at
$y_{W\!W}\sim 0$, the photon-quark induced processes give a positive
contribution which in addition modifies the shape of the EW
corrections into a mexican hat shape with two degenerate positive
minima at $|y_{W\!W}| \sim 2$. The photon radiated processes give a
small contribution of less that $+0.5\%$ on the whole rapidity range
considered and the diphoton corrections are dominant for
$|y_{WW}|\leq 3\%$.

From the above discussion about the huge QCD corrections originating
from the gluon-quark induced processes, we may think of using a jet veto
(i.e. veto events with $p_{T,\text{jet}} > p_\text{veto}$)
to reduce those corrections. This issue has been studied in
Ref.~\cite{Nhung:2013jta} for
a similar process of $WWZ$ production at the LHC. There it is found that
using
a dynamic jet veto such as $p_\text{veto} = 
\text{Max}(M_{T,V},M_{T,V^\prime})/2$ with
$M_{T,V}=(p_{T,V}^2 + M_V^2)^{1/2}$ being the transverse mass reduces
significantly
the QCD corrections and gives a stable result. Using a fixed jet veto
such as $p_\text{veto} =
25$ GeV is not a good idea because it over-subtracts the QCD corrections
and creates large
negative corrections at large $p_{T,V}$. However, the price to pay is
that the theoretical
uncertainty gets larger for exclusive observables with jet veto, see
e.g. Refs.~\cite{Stewart:2011cf, Nhung:2013jta}.

\subsection{Discussion of the leading-logarithmic
  approximation\label{discussion}}

We have seen in the previous subsections that the leading-logarithmic
approximation gives a good explanation why the QCD (EW) corrections
arising from the gluon (photon)-quark induced processes are largest for the
$WZ$ case and smallest for the $ZZ$ case. In order to have more
insights into this hierarchy, we compare the QCD results at the same
value of $p_T \gg M_Z$ so that the double-logarithmic factors are
approximately equal and can be ignored. We have
\begin{align}
\fr{d\sigma^{qg \to ZZq}}{d\sigma^{\bar{q}q \to ZZ}} &\propto
2c^u_{ZZ}\fr{d\sigma_L^{ug\to Zu}}{d\sigma^{\bar{q}q \to ZZ}},\crn
\fr{d\sigma^{ug \to W^+Zd}}{d\sigma^{u\bar{d} \to W^+Z}} &\propto
c^u_{WZ}\fr{d\sigma_L^{ug\to Zu}}{d\sigma^{u\bar{d} \to W^+Z}},\crn
\fr{d\sigma^{qg \to W^+W^-q}}{d\sigma^{\bar{q}q \to W^+W^-}} &\propto
c^u_{WW}\fr{d\sigma_L^{ug\to Zu}}{d\sigma^{\bar{q}q \to W^+W^-}},
\end{align}
where we have used the fact that $c^d_{ZZ}d\sigma_L^{dg\to Zd}
\approx c^u_{ZZ}d\sigma_L^{ug\to Zu}$ which is a consequence of the
two properties: the $Z$ boson couples more strongly to the $d$ quarks
than to the $u$ quarks, but the $d$ PDF is smaller than the $u$
PDF. Thus, the hierarchy of the numerators is
$c^u_{WZ}:c^u_{WW}:2c^u_{ZZ} = 4.13:3.53:0.36$. We observe a clear
difference between the $WV$ ($V=Z,W$) channels and the $ZZ$ channel
due to the fact that the former includes trilinear gauge couplings,
and also because of  the fact that the quarks couple more strongly to
the $W$ than to the $Z$ bosons. The hierarchy of the denominators is
$d\sigma^{\bar{q}q \to W^+W^-}:d\sigma^{u\bar{d} \to
  W^+Z}:d\sigma^{\bar{q}q \to ZZ} = 4.54:1.35:1$. This hierarchy due
to different PDFs and different coupling strengths is more reduced
compared to the one of the numerators. To sum up, the hierarchy
observed in the QCD gluon-quark induced corrections comes from trilinear
gauge couplings, different coupling strengths, and different PDFs. For
the EW photon-quark induced corrections, a larger hierarchy occurs. This is
because, in addition to the above mechanisms, there is a new dynamical
enhancement in the numerators for the $WV$ channels, namely the
contributions $d\sigma_L^{u\gamma \to W^+d}$ and $d\sigma_{LT}^{uW^+
  \to W^+u}$ with a $t$-channel exchange of a gauge boson are from one
to two orders of magnitude higher than $d\sigma_L^{u\gamma \to Zu}$.

\begin{figure}[h]
  \centering
  \includegraphics[scale=0.5]{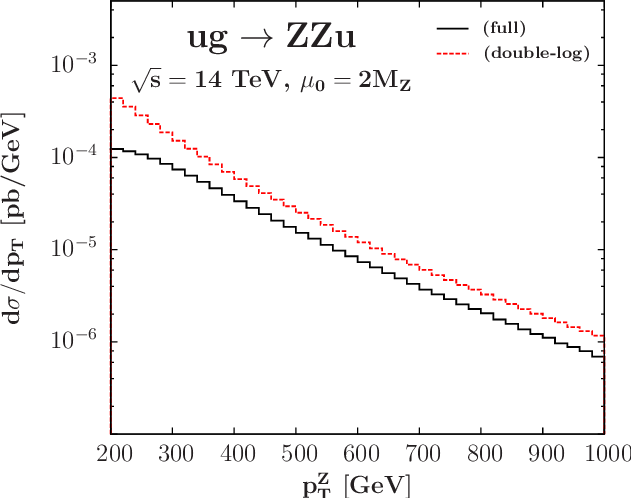}
  \hspace{2mm}
  \includegraphics[scale=0.5]{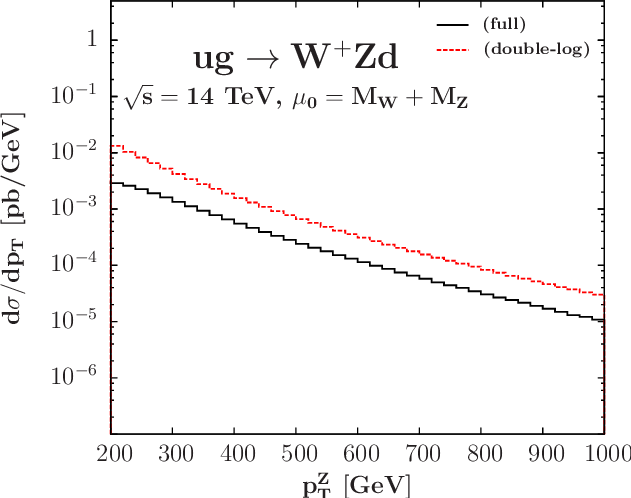}
  \hspace{2mm}  
  \includegraphics[scale=0.5]{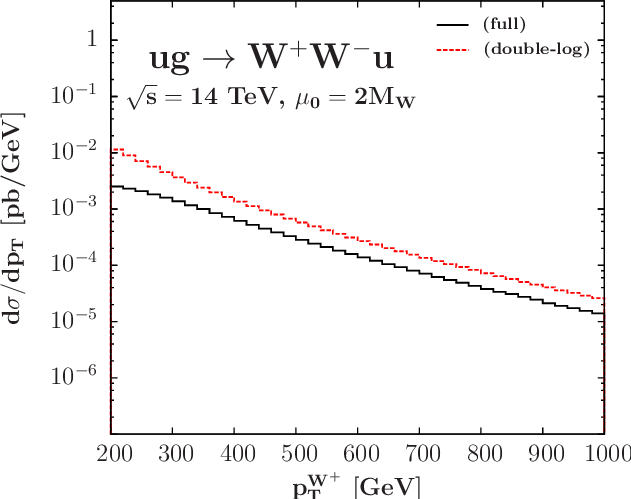}
  \caption{\small Comparison between the full results (plain black line) and
    the analytical leading-logarithmic approximation (dashed red line) for
    gluon-quark induced processes in $ZZ$ (left), $W^+Z$ (middle) and $WW$
    (right) $p_T$ differential distributions.}
  \label{fig:QCD_double_log}
\end{figure}
\begin{figure}[h]
  \centering
  \includegraphics[scale=0.5]{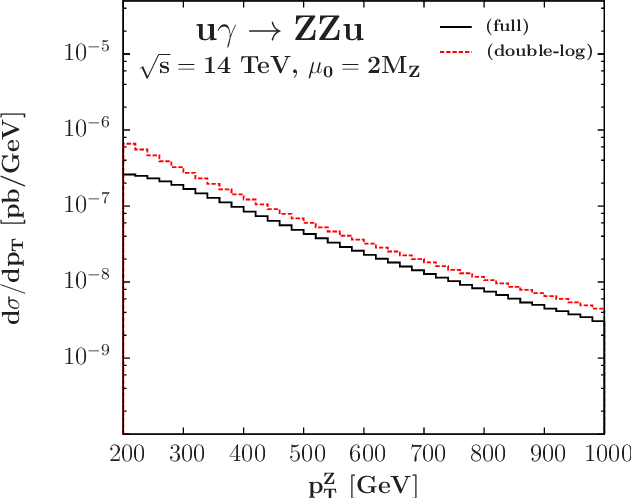}
  \hspace{2mm}
  \includegraphics[scale=0.5]{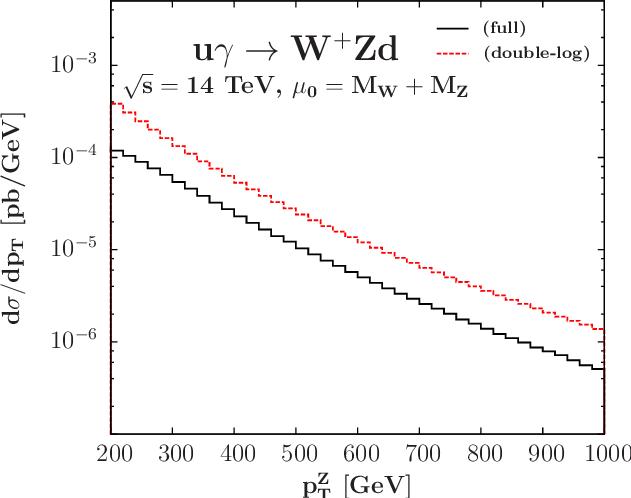}
  \hspace{2mm}
  \includegraphics[scale=0.5]{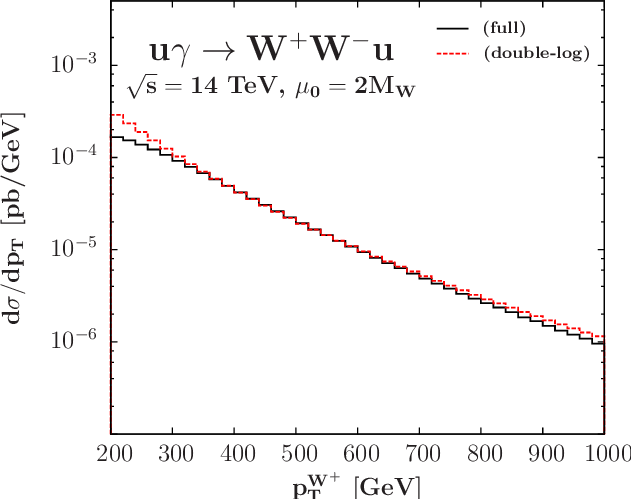}
  \caption{\small Same as \fig{fig:QCD_double_log} but for photon-quark induced processes.}
  \label{fig:EW_double_log}
\end{figure}
We observe, however, that the leading-logarithmic approximation can
differ from the full result by up to a factor of two in the case of
$WZ$ production.  This could trigger concerns about our explanation of
the hierarchy between the different diboson production channels. In
order to validate our calculation we display in this subsection the
comparison between the full result and the analytical approximation at
high transverse momentum. In \fig{fig:QCD_double_log} are displayed
the results for the gluon-quark induced processes. We only show the full
results without the subtraction term and we have used a transverse
momentum cut $p^q_T > 200$~GeV both in the full result and in the
analytical approximation. We have checked that the IR subtraction
terms included in the gluon-quark induced and photon-quark induced
contributions in the previous subsections are very small at large
transverse momentum ($p_T>500$~GeV). In all cases the analytical
approximation is larger than the full results. 
The difference shows the impact of single-logarithm terms that are
particularly sizable in the $WZ$ channel. To have a numerical feeling
for the difference between double and single logarithmic
corrections, assuming that the correction pre-factors are the same, we
get, as a ratio between double and single logarithmic contributions,
$\log(p_T^2/M_W^2)$ about $5$ at $p_T=1$~TeV and $20$ at
$10^3$~TeV. We have checked that the double-logarithmic contributions
almost coincide with the full results at $p_T=10^3$~TeV at a
hypothetical super hadron collider.

The same comparison has been done in the case of photon-quark induced
processes. This is displayed in \fig{fig:EW_double_log} below. In all
channels the agreement is better in the case of EW corrections
compared to QCD corrections, in particular in the $WW$ channel where
the agreement is almost perfect already at $p_T\simeq 700$ GeV.

\subsection{Contributions from third-generation external
  quarks\label{bquark}}

In this section we discuss the contributions with a $b$ quark in the
initial state. This is of relevance only for the $WW$ and $ZZ$
cases  because final states with a $t$ quark occurring in the $WZ$ case
lead to different experimental signatures and are therefore excluded. This
contribution is suppressed in the $ZZ$ case and at LO in the $W^+W^-$
case due to small $b$ PDF. For the $W^+W^-$ final state, there is a
new enhancement mechanism in the hard amplitude with an intermediate
on-shell top quark, as noticed in \bib{WW-EW-Kuhn}.
\begin{figure}[h]
  \centering
  \includegraphics[scale=0.53]{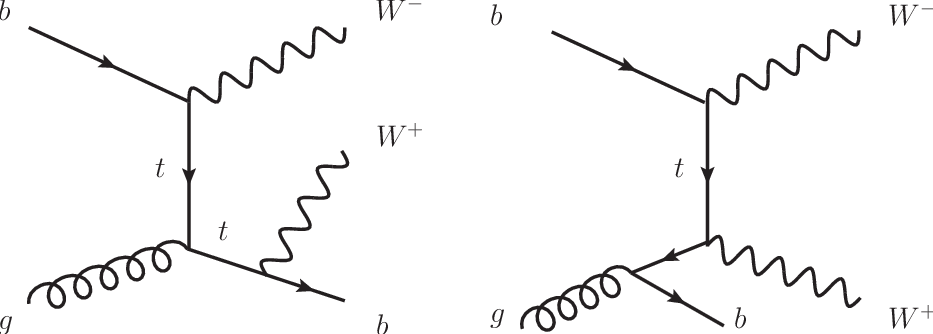}
  \caption{\small Representative $t$-channel diagrams for $b g\to W^+
    W^- b$ production cross section.}
  \label{fig:bquark_diagram}
\end{figure}
This occurs at NLO as shown in \fig{fig:bquark_diagram} for the
gluon-quark induced process $bg \to W^+ W^- b$. Only representative
$t$-channel diagrams are displayed but the discussion is similar for
the $s$-channel diagrams. The left-hand side of
\fig{fig:bquark_diagram} is a typical single top production channel
diagram, $b g\to t W^-$ followed by the decay $t\to W^+ b$ with
branching ratio close to one. The right-hand side diagram is
essentially $b\bar{b} \to W^+ W^-$ with one initial $b$ quark
originated from the gluon splitting. The large $tW$ production mode,
being a part of the singe-top background, should be excluded and our
main concern is the interference between the two mechanisms. We have
estimated this effect and found that, for $\sqrt{s}=14$~TeV, the
correction is at the per mille level compared to the full NLO total
cross section. We therefore conclude that, after the subtraction of
the single-top contribution, the contribution from processes with
initial $b$ quarks can be neglected.


\section{Total cross sections and theoretical
  uncertainties\label{section-4}}

In order to make a comparison with experimental results, a thorough
assessment of theoretical uncertainties affecting the central
predictions of the total cross sections is needed. In this section we
will consider three sources of uncertainties: the scale uncertainty
which is an estimate of the missing higher-order terms in the
perturbative calculation, the uncertainty related to the parton
distribution functions and the fitted value of the strong coupling
constant $\alpha_s(M_Z^2)$ and the parametric uncertainties related to
the experimental errors on $W$ and $Z$ masses, $M_W=(80.385 \pm
0.015)$ GeV and $M_Z = (91.1876 \pm 0.0021)$ GeV.

As for the last source of uncertainties mentioned above, we have
checked that it actually does not affect our predictions more than
$\pm 0.2\%$ at all c.m. energies in all diboson
channels. We will then ignore the parametric errors in the final
combination of the theoretical uncertainties.

In order to include the full NLO QCD+EW effects using different PDF
sets, the calculation is done in two steps. First, using
{\texttt{MRST2004QED}} PDF set~\cite{PDF-QED} which includes the
photon PDF as described in \sect{sect:cal_details}, we calculate the 
EW correction factor as
\begin{align}
\delta_{\rm EW} = \frac{\sigma^{\rm QCD+EW,\, MRSTQED}}{\sigma^{\rm
    QCD,\, MRSTQED}}\, ,
\end{align}
where the numerator is the full result including also the gluon fusion 
and the $\gamma\gamma$ channel at NLO (for the $WW$ case), 
while the denominator is without 
the NLO EW corrections and the $\gamma\gamma$ contribution. 
We then calculate the full NLO QCD correction using another PDF set,
e.g. {\texttt{MSTW2008}} set~\cite{PDF-MSTW}, and rescale it with 
the above EW correction factor to get 
\begin{align}
\sigma^{\rm QCD+EW}_{\rm MSTW} = \delta_{\rm EW} \times \sigma^{\rm
  QCD}_{\rm MSTW}\, .
\end{align}
We use the same parameter setup as in \sect{section-3}, save the value of
$\alpha_s(M_Z^2)$ that is adapted according to the PDF set used. In
the case of {\texttt{MSTW2008}} PDF set that we will use as our
default set, we will use $\alpha_s^{\rm NLO}(M_Z^2) = 0.12018$ and
$\alpha_s^{\rm NNLO}(M_Z^2)=0.11707$ as central values. The NNLO value
(and running) of the strong coupling constant is used for $gg\to
WW,ZZ$ subprocesses as well as the NNLO gluon PDF when
available. Otherwise the strong coupling constant gets its NLO value
and the running is also evaluated at NLO.

We have found that EW corrections are negligible in the $WZ$
channels. In the case of $WW$ channel we have found $\delta_{\rm
  EW}^{WW} \sim +1.5\%$ mainly due to the $\gamma\gamma$ contribution 
while in the case of $ZZ$ channel the
correction is negative and sizable, $\delta_{\rm EW}^{ZZ} \sim
-3\%$ due to the EW virtual corrections, 
with nearly no energy dependence for both production
channels. Note that this EW factor is the size of the EW corrections
on top of the full NLO QCD predictions.

\subsection{Scale uncertainty}

As the calculation is done in the perturbative framework, the
theoretical cross sections depend on two unphysical scales: the
renormalization scale $\mu_R$ that comes from the running of
$\alpha_s$, and the factorization scale $\mu_F$ that comes from the
convolution of the perturbative partonic cross section with the
non-perturbative parton distribution functions. The variation of the
cross sections with respect to these two scales can be viewed as an
estimate of the missing higher-order corrections and this is the first
uncertainty that is considered in this paper. We choose the interval
\begin{align}
\frac12 \mu_0 \leq \mu_R=\mu_F \leq 2\,\mu_0 \;,
\end{align}
where $\mu_0$ is the central scale for the process under study and has
been defined in the previous section.

\begin{figure}[ht!]
\begin{center}
\includegraphics[scale=0.75]{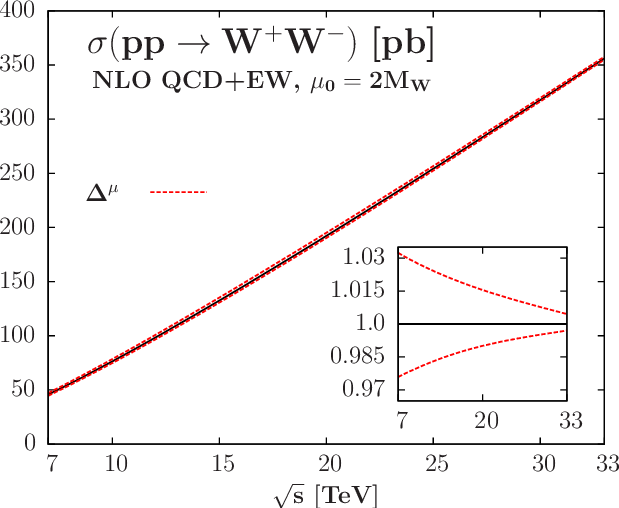}
\hspace{6mm}
\includegraphics[scale=0.75]{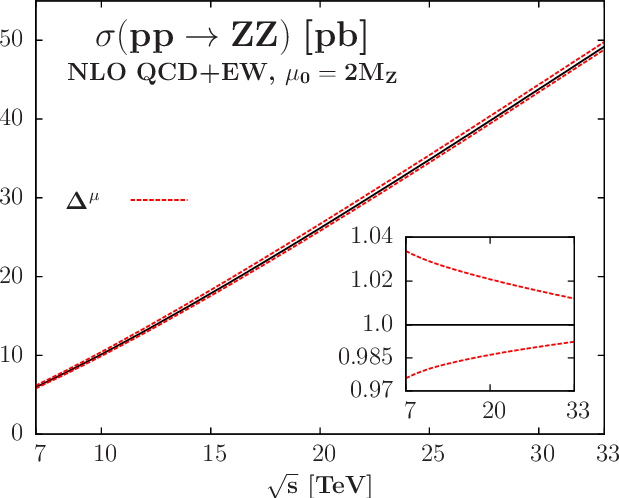}

\vspace{3mm}
\includegraphics[scale=0.75]{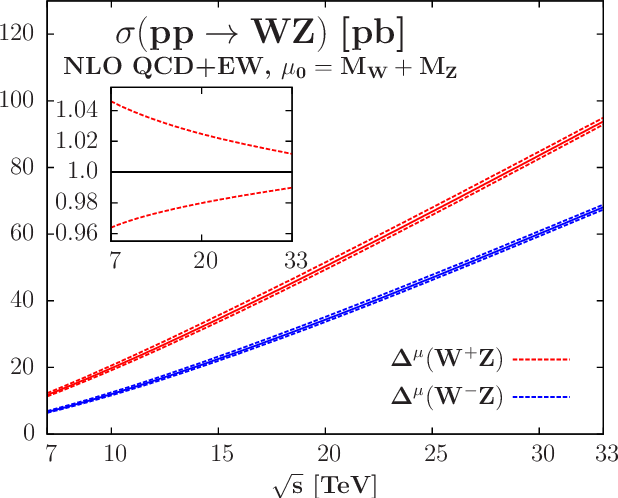}
\end{center}
\it{\vspace{-4mm}\caption{\small Scale uncertainty for a scale variation in
    the interval $\frac12 \mu_0 \leq \mu_R=\mu_F \leq 2\mu_0$ in
    $\sigma(pp\to WW, ZZ, WZ)$ (in pb) at the LHC as a
    function of $\sqrt{s}$ (in TeV). Upper left: $WW$ cross
    section. Upper right: $ZZ$ cross section. Lower: $W^+Z$ and $W^-Z$
    cross sections. In the inserts the relative deviation from the
    central cross section obtained with $\mu_R=\mu_F = \mu_0 =
    M_{V_1}+M_{V_2}$ is shown, $V_i = W/Z$.\label{VV-scale-error}}}
\end{figure}

As can be seen in \fig{VV-scale-error}, the scale uncertainty is
limited in the different gauge boson pair production channels: we
obtain $\sim +3\% / -2.5\%$ at 7 TeV in $WW$ and $ZZ$ channels,
slightly more in $WZ$ channels with $\sim +5\%/-4\%$ at 7 TeV. It
then reduces down to less than $1\%$ at 33 TeV in all diboson production
channels.

\subsection{PDF+\texorpdfstring{$\alpha_s$}{alphaS} uncertainty}

\begin{figure}[ht!]
\begin{center}
\includegraphics[scale=0.75]{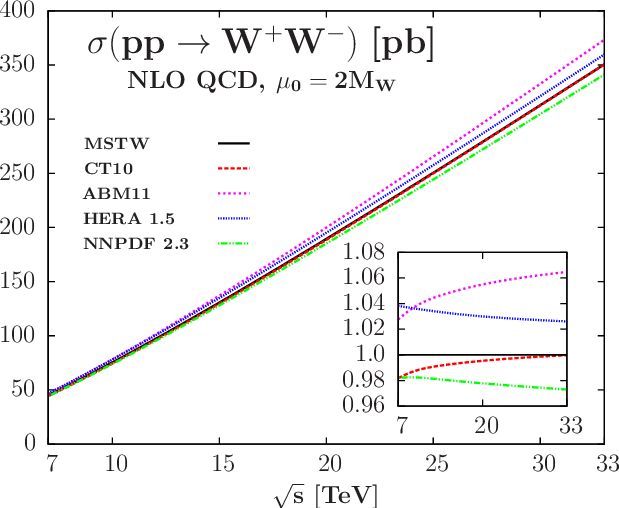}
\hspace{6mm}
\includegraphics[scale=0.75]{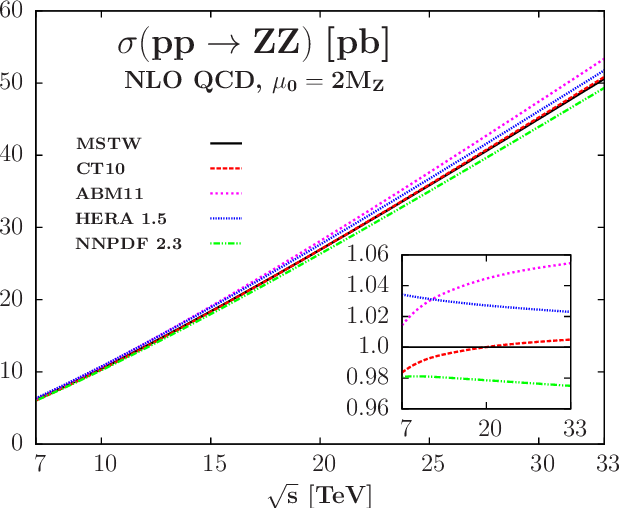}

\vspace{3mm}
\includegraphics[scale=0.75]{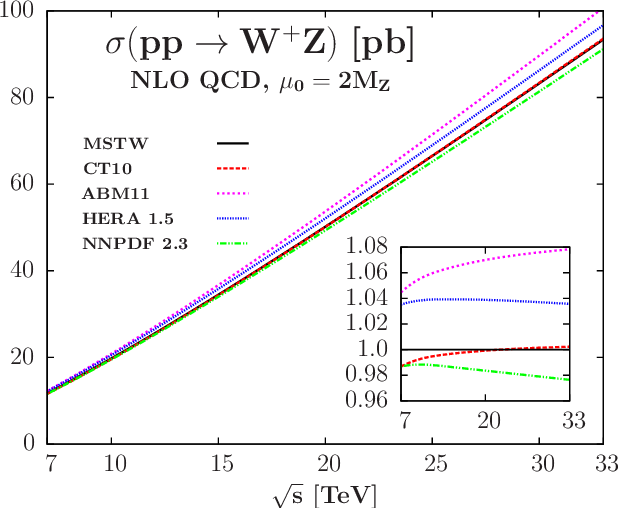}
\hspace{6mm}
\includegraphics[scale=0.75]{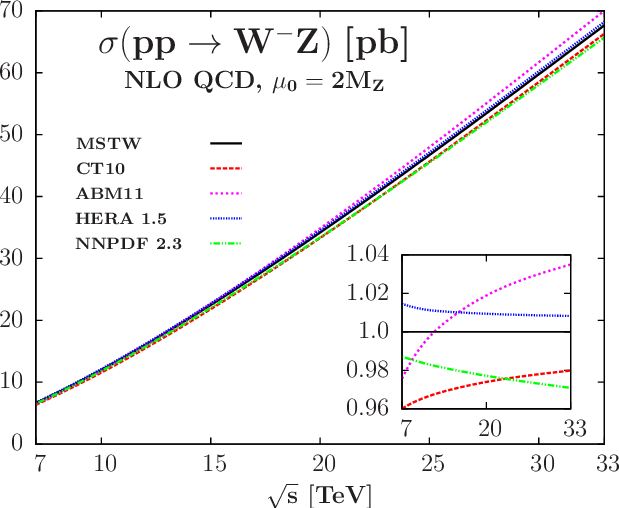}
\end{center}
\it{\vspace{-4mm}\caption{\small Total cross sections (in pb) as a function
    of the c.m. energy (in TeV) for $WW$ (upper left), $ZZ$ (upper
    right), $W^+Z$ (lower left) and $W^-Z$ (lower right) production
    channels at the LHC when using different PDF sets. In the inserts
    the relative deviation from the central cross section obtained
    with {\texttt{MSTW2008}} PDF set is shown.\label{VV-pdf-diff}}}
\end{figure}

The other main source of theoretical uncertainty is the
parametrization of the parton distribution functions (PDF). Indeed the
calculation of an hadronic cross section is done in two parts: first
one calculates (in a perturbative framework) the partonic cross
section, then the result is convoluted with the non-perturbative
parton distribution functions that are the probability distribution of
extracting from the proton a given parton with a momentum fraction $x$
of the initial proton. The PDFs are fitted quantities on experimental
data sets and that leads to uncertainties in the theoretical
calculation. There are many different sets on the market, depending on
the choice of the parametrization, the set of input parameters used,
the running of the parameters, etc. One way to quantify the pure
theoretical uncertainties induced by these differences is to compare
the predictions obtained with the various sets, such as
{\texttt{MSTW2008}}~\cite{PDF-MSTW}, {\texttt{CT10}}~\cite{PDF-CT10},
{\texttt{ABM11}}~\cite{PDF-ABKM}, {\texttt{HERA}}~\cite{PDF-HERA} or
{\texttt{NNPDF 2.3}}~\cite{PDF-NNPDF}. Each PDF collaboration uses
different experimental and theoretical assumptions, e.g. which data to
be used to build the fit, heavy--flavor scheme, running of the
parameters, etc. In addition, the value of the strong coupling
constant $\alpha_s(M_Z^2)$ is also fitted together with the PDF
sets. The MSTW Collaboration central value is $\alpha_s(M_Z^2)=0.12018$ at
NLO, while CT10 uses $\alpha_s(M_Z^2)=0.118$, the ABM11 central value is
$\alpha_s(M_Z^2)=0.11797$, HERA 1.5 uses $\alpha_s(M_Z^2)=0.1176$ and
the NNPDF 2.3 central value is $\alpha_s(M_Z^2)=0.117$. We display in
\fig{VV-pdf-diff} the total cross sections for $WW$, $ZZ$ and
$WZ$ production at the LHC when using these different PDF sets. Only
best-fit sets are used and this exemplifies the sizable differences
between the various predictions, up to $+6\%$ in $WW$ and $ZZ$ channel
from ABM11 PDF set with respect to MSTW PDF set and even more in
$W^+Z$ with a $+8\%$ increase. The lower values are generally obtained
with NNPDF PDF set which deviates between $-2\%$ and $-4\%$ from the
prediction obtained with MSTW PDF set. CT10 predictions tend to be
closer to MSTW predictions except for the case of $W^-Z$ production
where CT10 prediction is lower than MSTW prediction by $2\%$ to
$4\%$. The numbers are also given in
Tables~\ref{table:WW-allpdfs},~\ref{table:ZZ-allpdfs},~\ref{table:WplusZ-allpdfs}
and \ref{table:WminusZ-allpdfs}. Note that we have only displayed
numbers for NLO QCD cross sections as the impact of EW corrections is
limited and cancel out in our scheme when comparing with MSTW
predictions.

\begin{table}[ht!]
 \renewcommand{\arraystretch}{1.3}
  \begin{center}
   \small
\begin{tabular}{|c|ccccc|}\hline
      $\sqrt{s}$ [TeV] & MSTW [pb] & CT10 [pb] & ABM11 [pb] & HERA 1.5
      [pb] & NNPDF 2.3 [pb] \\\hline
      $7$ & $45.06$ & $44.24$ & $46.30$ & $46.78$ & $44.25$ \\\hline
      $8$ & $54.88$ & $54.03$ & $56.64$ & $56.94$ & $53.92$ \\\hline
      $14$ & $119.28$ & $118.32$ & $124.90$ & $123.22$ & $116.97$
      \\\hline
      $33$ & $350.63$ & $350.57$ & $373.32$ & $359.78$ & $341.16$
      \\\hline
\end{tabular}
\it{\caption[]{\small The total $W$ boson pair production cross
    section at NLO in QCD at the LHC (in pb) for given c.m. energies
    (in TeV) at the central scale $\mu_F=\mu_R=2 M_W$, when using
    MSTW, CT10, ABM11, HERA 1.5 and NNPDF 2.3 PDF sets.}
  \label{table:WW-allpdfs}}
\end{center}
\vspace*{-0.1cm}
\end{table}

\begin{table}[ht!]
 \renewcommand{\arraystretch}{1.3}
  \begin{center}
   \small
\begin{tabular}{|c|ccccc|}\hline
      $\sqrt{s}$ [TeV] & MSTW [pb] & CT10 [pb] & ABM11 [pb] & HERA 1.5
      [pb] & NNPDF 2.3 [pb] \\\hline
      $7$ & $6.13$ & $6.03$ & $6.22$ & $6.34$ & $6.01$ \\\hline
      $8$ & $7.52$ & $7.42$ & $7.66$ & $7.77$ & $7.37$ \\\hline
      $14$ & $16.73$ & $16.66$ & $17.33$ & $17.23$ & $16.39$ \\\hline
      $33$ & $50.61$ & $50.86$ & $53.37$ & $51.77$ & $49.34$ \\\hline
\end{tabular}
\it{\caption[]{\small Same as \protect\tab{table:WW-allpdfs} but for
    $ZZ$ case at the central scale $\mu_F=\mu_R=2 M_Z$.}
  \label{table:ZZ-allpdfs}}
\end{center}
\vspace*{-0.1cm}
\end{table}

\begin{table}[ht!]
 \renewcommand{\arraystretch}{1.3}
  \begin{center}
   \small
\begin{tabular}{|c|ccccc|}\hline
      $\sqrt{s}$ [TeV] & MSTW [pb] & CT10 [pb] & ABM11 [pb] & HERA 1.5
      [pb] & NNPDF 2.3 [pb] \\\hline
      $7$ & $11.75$ & $11.59$ & $12.27$ & $12.16$ & $11.59$ \\\hline
      $8$ & $14.34$ & $14.19$ & $15.04$ & $14.87$ & $14.17$ \\\hline
      $14$ & $31.44$ & $31.32$ & $33.42$ & $32.68$ & $31.03$ \\\hline
      $33$ & $93.38$ & $93.60$ & $100.70$ & $96.71$ & $91.18$ \\\hline
\end{tabular}
\it{\caption[]{\small Same as \protect\tab{table:WW-allpdfs} but for
    $W^+ Z$ case at the central scale $\mu_F=\mu_R=M_W+M_Z$.}
  \label{table:WplusZ-allpdfs}}
\end{center}
\vspace*{-0.3cm}
\end{table}

\begin{table}[ht!]
 \renewcommand{\arraystretch}{1.3}
  \begin{center}
   \small
\begin{tabular}{|c|ccccc|}\hline
      $\sqrt{s}$ [TeV] & MSTW [pb] & CT10 [pb] & ABM11 [pb] & HERA 1.5
      [pb] & NNPDF 2.3 [pb] \\\hline
      $7$ & $6.60$ & $6.34$ & $6.44$ & $6.70$ & $6.52$ \\\hline
      $8$ & $8.31$ & $7.99$ & $8.16$ & $8.42$ & $8.19$ \\\hline
      $14$ & $20.26$ & $19.64$ & $20.38$ & $20.47$ & $19.88$ \\\hline
      $33$ & $67.67$ & $66.31$ & $70.04$ & $68.23$ & $65.70$ \\\hline
\end{tabular}
\it{\caption[]{\small Same as \protect\tab{table:WW-allpdfs} but for
    $W^- Z$ case at the central scale $\mu_F=\mu_R=M_W+M_Z$.}
  \label{table:WminusZ-allpdfs}}
\end{center}
\vspace*{-0.1cm}
\end{table}

Besides the differences between the sets, there are experimental
uncertainties associated with the experimental data used to build the
fit. MSTW, CT10, HERA and ABM Collaborations use the Hessian method to
build additional sets next to the best-fit PDF to account for the
experimental uncertainties in the data used to build the distribution
functions. Additional $2 N_{\rm PDF}$ sets are built from the $\pm
1\sigma$ variation around the minimal $\chi^2$ of all $N_{\rm PDF}$
parameters that enter the fit, the tolerance interval depending on the
collaboration. Note that the NNPDF Collaboration uses an alternative
method to build the additional sets based on Monte-Carlo
replicas. Using the $90\%$ CL error PDF sets provided by the MSTW
Collaboration a PDF error of about $\sim +3.5\% / -3.0\%$ is obtained
for $\sqrt{s} = 7$ TeV in the $WW$ and $ZZ$ channels, slightly more in
the $WZ$ channels with an error of $\sim \pm 4.0\%$. The uncertainty
reduces down to $\sim +3.0\% / -2.5\%$ ($\sim \pm 3\%$) at
$\sqrt{s}=33$ TeV in the $WW$ and $ZZ$ channels ($WZ$ channels).

On top of the pure PDF uncertainty, the value of the strong coupling
constant $\alpha_s$ induces also an uncertainty in the theoretical
prediction of the hadronic cross sections. Even if this will not be a
dominant effect in diboson production as the three different channels
are purely electroweak processes at leading order, the impact of the
uncertainty on $\alpha_s(M_Z^2)$ is not negligible as the QCD
corrections are large. The MSTW Collaboration provides
additional PDF sets such that the combined PDF+$\alpha_s$
uncertainties can be evaluated in a consistent way taking into account
the correlation between the PDF and $\alpha_s$~\cite{PDF-MSTW-as}. The
fitted value of $\alpha_s(M_Z^2)$ is then:
\begin{align}
  \alpha_s^{\rm NLO}(M_Z^2)=0.12018^{+0.00122}_{-0.00151}\text{(at
    68$\%$ CL) or }^{+0.00317}_{-0.00386}\text{ (at 90$\%$ CL)}\;,\nonumber\\
  \alpha_s^{\rm NNLO}(M_Z^2)=0.11707^{+0.00141}_{-0.00135}\text{(at
    68$\%$ CL) or }^{+0.00337}_{-0.00342}\text{ (at 90$\%$ CL)}\;,
\label{error_alphaS_exp}
\end{align}
and with the $90\%$ CL error PDF sets we obtain a PDF$+\alpha_s$ error
that is slightly larger that the pure PDF uncertainty in the three
different diboson channels: $\sim +4.2\% / -3.5\%$ in the $WW$ and
$ZZ$ channels, $\sim +4.5\% / -4.0\%$ in the $WZ$ channels, all at
$\sqrt{s} = 7$ TeV. It then reduces down to $\sim \pm 4.0\%$ in the
three diboson channels at $\sqrt{s}=33$ TeV. Note that $W^+Z$ and
$W^-Z$ channels have similar uncertainties, the difference being up to
$\sim 0.5\%$ at 7 TeV in the lower deviation.

\begin{figure}[ht!]
\begin{center}
\includegraphics[scale=0.75]{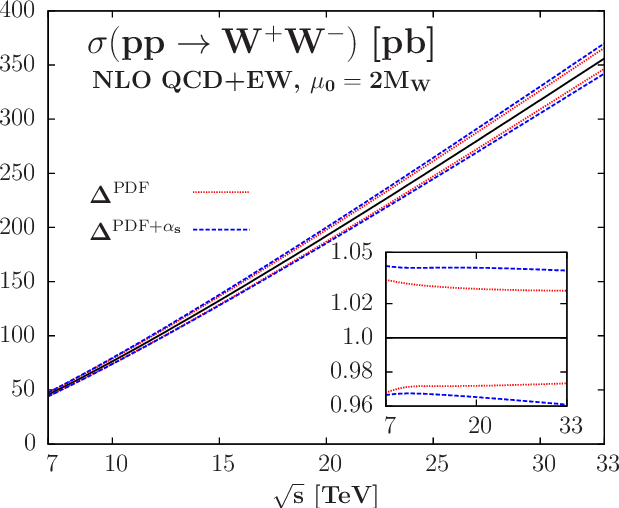}
\hspace{6mm}
\includegraphics[scale=0.75]{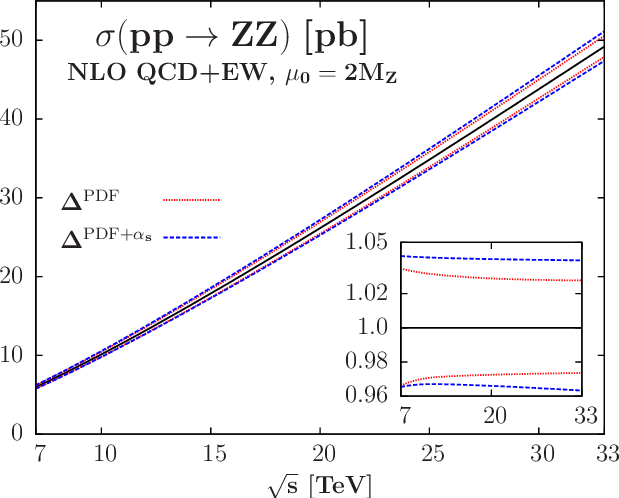}

\vspace{3mm}
\includegraphics[scale=0.75]{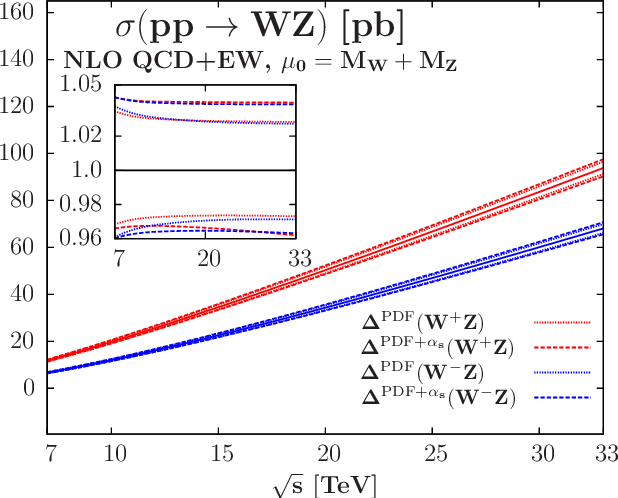}
\end{center}
\it{\vspace{-4mm}\caption{\small PDF and PDF$+\alpha_s$ uncertainties using
    {\texttt{MSTW2008}} PDF set in $\sigma(pp\to WW, ZZ, WZ)$ (in pb)
    at the LHC as a function of the c.m. energy $\sqrt{s}$ (in
    TeV). Upper left: $WW$ cross section. Upper right: $ZZ$ cross
    section. Lower: $W^+Z$ and $W^-Z$ cross sections. In the inserts
    the relative deviation from the central cross section is
    shown.\label{VV-pdf-error}}}
\end{figure}

As discussed in Ref.~\cite{HH-Julien} in the case of Higgs pair
production, a theoretical uncertainty on $\alpha_s$ could be
considered, stemming from scale variation or ambiguities in the heavy
flavor scheme definition. The MSTW Collaboration estimates this
uncertainty for $\alpha_s$ to be $\Delta^{\rm th} \alpha_s(M_Z^2) =
\pm 0.003$ at NLO and $\Delta^{\rm th} \alpha_s(M_Z^2) = \pm 0.002$ at
NNLO~\cite{PDF-MSTW-as}. However, this uncertainty is already included
in the scale uncertainty on the input data sets included in the global
fit of the PDF and therefore has already been accounted for by the
PDF+$\alpha_s$ error. We will not consider it separately and our final
PDF+$\alpha_s$ uncertainty will be the {\texttt{MSTW2008}}
PDF$+\alpha_s$ uncertainty, exemplified in \fig{VV-pdf-error}. It can
be noted that in general it accounts for the differences between the
various PDF sets seen in \fig{VV-pdf-diff}.

\subsection{Total uncertainty and comparison with experimental cross
  sections}

\begin{figure}[ht!]
\begin{center}
\includegraphics[scale=0.7]{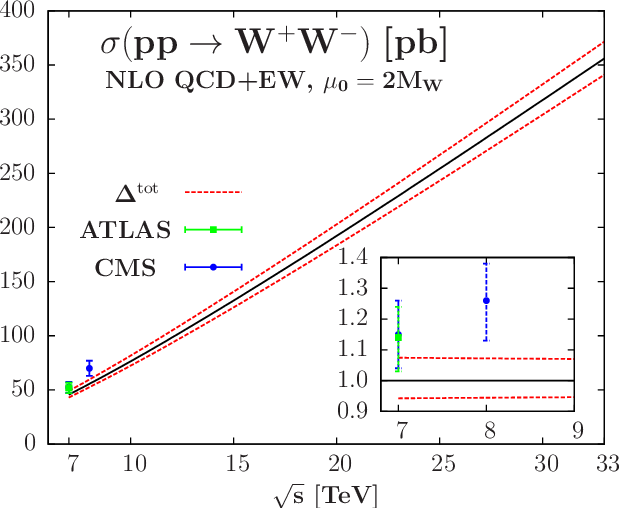}
\hspace{6mm}
\includegraphics[scale=0.7]{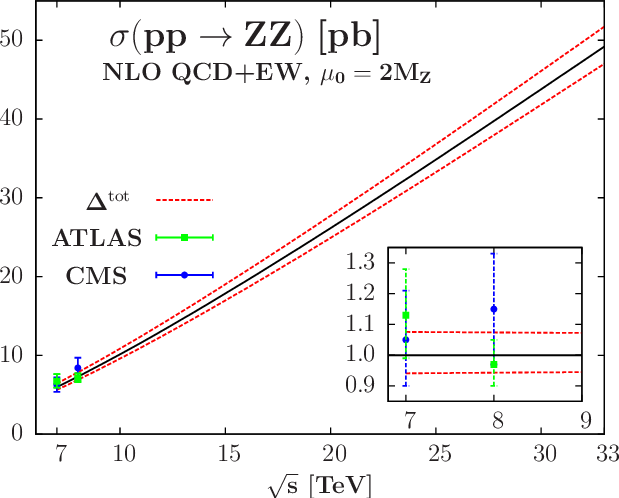}

\vspace{3mm}
\includegraphics[scale=0.7]{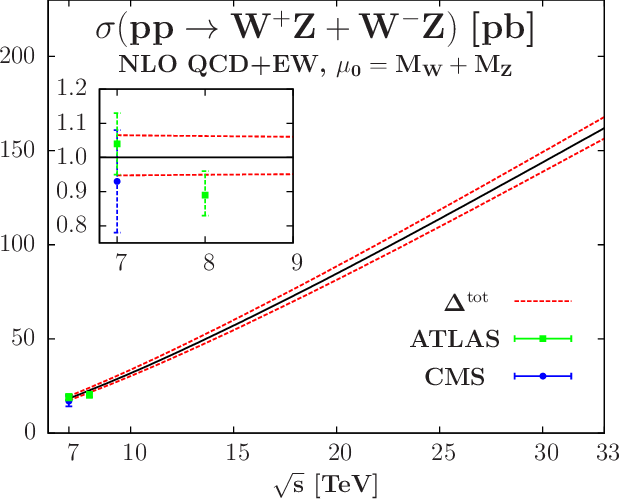}
\end{center}
\it{\vspace{-4mm}\caption{\small The NLO QCD+EW total cross section
    (black/full, in pb) of the processes $pp\to WW$ (upper left),
    $pp\to ZZ$ (upper right) and $pp\to W^+Z+W^-Z$ (lower) at the
    LHC as a function of the c.m. energy $\sqrt{s}$ (in TeV) including
    the total theoretical uncertainty (red/dashed) as discussed in the
    text. The insert shows the relative deviation from the central
    cross sections, and the experimental data points are also
    displayed on the main figures.\label{VV-total-error}}}
\end{figure}

Before comparing with experimental data on diboson production, we
combine the errors to obtain our prediction for the total theoretical
uncertainty. Following the LHC Higgs Cross Section Working
Group~\cite{LHCXS}, we do not use a quadratic addition that would be
too optimistic and simply add linearly scale and PDF$+\alpha_s$
uncertainties. We do not use the alternative combination presented in
Ref.~\cite{Julien-errors}  in the case of Higgs production, the scale
uncertainty being too limited to have sizable effects on the
calculating of the PDF$+\alpha_s$ uncertainty. We obtain sizable
uncertainties in the different diboson production channels, ranging
from $\sim +8\% / -6\%$ at 7 TeV in $WW$ and $ZZ$ channels, $\sim
+9\% / -8\%$ in $WZ$ channels, down to $\sim +4-5\% / -4\%$ at 33 TeV
in $WW$ and $ZZ$ channels, $\sim \pm 5\%$ in $WZ$ channels. This is
displayed in \fig{VV-total-error} and also detailed in
Tables~\ref{table:ww-total-errors}, \ref{table:zz-total-errors} and
\ref{table:wz-total-errors}.

\begin{table}[ht!]
 \renewcommand{\arraystretch}{1.3}
  \begin{center}
   \small
\begin{tabular}{|c|ccccc|}\hline
      $\sqrt{s}$ [TeV] & $\sigma^{\rm NLO}_{WW}$ [pb] & Scale
      [\%] & PDF [\%] & PDF+$\alpha_s$ [\%] & Total [\%] \\\hline
      $7$ & $45.65$ & ${+3.2}\;\;\;{-2.4}$ & ${+3.4}\;\;\;{-3.2}$ &
      ${+4.2}\;\;\;{-3.3}$ & ${+7.4}\;\;\;{-5.8}$
      \\\hline
      $8$ & $55.61$ & ${+3.1}\;\;\;{-2.3}$ & ${+3.3}\;\;\;{-3.1}$ &
      ${+4.2}\;\;\;{-3.3}$ & ${+7.2}\;\;\;{-5.6}$
      \\\hline
      $14$ & $120.96$ & ${+2.2}\;\;\;{-1.5}$ & ${+3.0}\;\;\;{-2.8}$ &
      ${+4.1}\;\;\;{-3.3}$ & ${+6.3}\;\;\;{-4.8}$
      \\\hline
      $33$ & $356.02$ & ${+0.5}\;\;\;{-0.3}$ & ${+2.8}\;\;\;{-2.7}$ &
      ${+3.9}\;\;\;{-3.9}$ & ${+4.4}\;\;\;{-4.2}$
      \\\hline
\end{tabular}
\it{\caption[]{\small The total $W$ boson pair production
    cross section at NLO in QCD+EW at the LHC (in pb) for given
    c.m. energies (in TeV) at the central scale $\mu_F=\mu_R=2
    M_W$. The corresponding deviations due to the theoretical
    uncertainties from the various sources discussed are shown as well
    as the total uncertainty when all errors are added linearly.}
  \label{table:ww-total-errors}}
\end{center}
\vspace*{-0.1cm}
\end{table}

\begin{table}[ht!]
 \renewcommand{\arraystretch}{1.3}
  \begin{center}
   \small
\begin{tabular}{|c|ccccc|}\hline
      $\sqrt{s}$ [TeV] & $\sigma^{\rm NLO}_{ZZ}$ [pb] & Scale
      [\%] & PDF [\%] & PDF+$\alpha_s$ [\%] & Total [\%] \\\hline
      $7$ & $5.95$ & ${+3.4}\;\;\;{-2.4}$ & ${+3.5}\;\;\;{-3.4}$ &
      ${+4.2}\;\;\;{-3.5}$ & ${+7.6}\;\;\;{-5.9}$
      \\\hline
      $8$ & $7.30$ & ${+3.2}\;\;\;{-2.3}$ & ${+3.4}\;\;\;{-3.2}$ &
      ${+4.2}\;\;\;{-3.4}$ & ${+7.4}\;\;\;{-5.7}$
      \\\hline
      $14$ & $16.24$ & ${+2.6}\;\;\;{-1.7}$ & ${+3.0}\;\;\;{-2.8}$ &
      ${+4.1}\;\;\;{-3.3}$ & ${+6.6}\;\;\;{-5.0}$
      \\\hline
      $33$ & $49.20$ & ${+1.2}\;\;\;{-0.8}$ & ${+2.8}\;\;\;{-2.6}$ &
      ${+3.9}\;\;\;{-3.7}$ & ${+5.1}\;\;\;{-4.4}$
      \\\hline
\end{tabular}
\it{\caption[]{\small Same as \protect\tab{table:ww-total-errors} but
    for $ZZ$ case at the central scale $\mu_F=\mu_R=2 M_Z$.}
  \label{table:zz-total-errors}}
\end{center}
\vspace*{-0.1cm}
\end{table}

\begin{table}[ht!]
 \renewcommand{\arraystretch}{1.3}
  \begin{center}
   \small
\begin{tabular}{|c|ccccc|}\hline
      $\sqrt{s}$ [TeV] & $\sigma^{\rm NLO}_{W^+Z}$ [pb] & Scale [\%] &
      PDF [\%] & PDF+$\alpha_s$ [\%] & Total [\%] \\\hline
      $7$ & $11.73$ & ${+4.6}\;\;\;{-3.6}$ & ${+3.5}\;\;\;{-3.2}$ &
      ${+4.3}\;\;\;{-3.4}$ & ${+8.9}\;\;\;{-7.0}$
      \\\hline
      $8$ & $14.33$ & ${+4.4}\;\;\;{-3.4}$ & ${+3.3}\;\;\;{-3.0}$ &
      ${+4.2}\;\;\;{-3.3}$ & ${+8.6}\;\;\;{-6.8}$
      \\\hline
      $14$ & $31.49$ & ${+3.3}\;\;\;{-2.6}$ & ${+3.0}\;\;\;{-2.7}$ &
      ${+4.0}\;\;\;{-3.3}$ & ${+7.3}\;\;\;{-5.9}$
      \\\hline
      $33$ & $93.84$ & ${+1.2}\;\;\;{-1.0}$ & ${+2.8}\;\;\;{-2.7}$ &
      ${+4.0}\;\;\;{-3.8}$ & ${+5.1}\;\;\;{-4.8}$
      \\\hline
\end{tabular}\vspace{3mm}

\begin{tabular}{|c|ccccc|}\hline
      $\sqrt{s}$ [TeV] & $\sigma^{\rm NLO}_{W^-Z}$ [pb] & Scale [\%] &
      PDF [\%] & PDF+$\alpha_s$ [\%] & Total [\%] \\\hline
      $7$ & $6.61$ & ${+4.6}\;\;\;{-3.6}$ & ${+3.7}\;\;\;{-3.9}$ &
      ${+4.3}\;\;\;{-4.0}$ & ${+8.9}\;\;\;{-7.6}$
      \\\hline
      $8$ & $8.32$ & ${+4.4}\;\;\;{-3.4}$ & ${+3.6}\;\;\;{-3.7}$ &
      ${+4.2}\;\;\;{-3.9}$ & ${+8.6}\;\;\;{-7.3}$
      \\\hline
      $14$ & $20.33$ & ${+3.3}\;\;\;{-2.6}$ & ${+3.0}\;\;\;{-3.1}$ &
      ${+3.9}\;\;\;{-3.6}$ & ${+7.3}\;\;\;{-6.2}$
      \\\hline
      $33$ & $68.09$ & ${+1.2}\;\;\;{-1.0}$ & ${+2.7}\;\;\;{-2.9}$ &
      ${+3.9}\;\;\;{-3.7}$ & ${+5.0}\;\;\;{-4.7}$
      \\\hline
\end{tabular}
\it{\caption[]{\small Same as \protect\tab{table:ww-total-errors} but
    for $W^+Z$ and $W^- Z$ cases at the central scale
    $\mu_F=\mu_R=M_W+M_Z$.}
  \label{table:wz-total-errors}}
\end{center}
\vspace*{-0.1cm}
\end{table}

We are now ready to compare with experimental results given by the ATLAS
and CMS Collaborations. The latest state-of-the-art total cross
section measurements at the LHC use a luminosity in the ATLAS
experiment of $4.6$ fb$^{-1}$ for the $WW$, $ZZ$ and $WZ$ measurements
at 7 TeV~\cite{ATLAS-WW-7tev, ATLAS-ZZ-7tev, ATLAS-WZ-7tev}, 13
fb$^{-1}$ for the $WZ$ measurement at 8 TeV~\cite{ATLAS-WZ-8tev} and
$20.3\pm 0.6$ fb$^{-1}$ for the $ZZ$ measurement at 8
TeV~\cite{ATLAS-ZZ-8tev}; the CMS experiment uses a luminosity of 1.1
fb$^{-1}$ for the $WZ$ measurement at 7 TeV~\cite{CMS-WZ-7tev}, 3.5
fb$^{-1}$ for the $WW$ measurement at 8 TeV~\cite{CMS-WW/ZZ-8tev},
4.92 fb$^{-1}$ for the $WW$ measurement at 7 TeV~\cite{CMS-WW-7tev}, 5
fb$^{-1}$ for the $ZZ$ measurement at 7 TeV~\cite{CMS-ZZ-7tev} and 5.3
fb$^{-1}$ for the $ZZ$ measurement at 8 TeV~\cite{CMS-WW/ZZ-8tev}.
\begin{table}[ht!]
\renewcommand{\arraystretch}{1.3}
\begin{center}
  \small
  \begin{tabular}{|c|ll|}\hline
    & ATLAS total cross section [pb] & CMS total cross section [pb]\\
    \hline
    $WW$ 7 TeV & $51.9 \pm 2.0 (\text{\footnotesize stat}) \pm 3.9
    (\text{\footnotesize syst}) \pm 2.0 (\text{\footnotesize lumi})$ &
    $52.4 \pm 2.0 (\text{\footnotesize stat}) \pm 4.5
    (\text{\footnotesize syst}) \pm 1.2 (\text{\footnotesize lumi})$\\
    $WW$ 8 TeV & & $69.9 \pm 2.8 (\text{\footnotesize stat}) \pm 5.6
    (\text{\footnotesize syst}) \pm 3.1 (\text{\footnotesize lumi})$
    \\ \hline
    $ZZ$ 7 TeV & $6.7 \pm 0.7 (\text{\footnotesize stat})
    ^{+0.4}_{-0.3} (\text{\footnotesize syst}) \pm 0.3
    (\text{\footnotesize lumi})$ & $6.24 ^{+0.86}_{-0.80}
    (\text{\footnotesize stat}) ^{+0.41}_{-0.32} (\text{\footnotesize
      syst}) \pm 0.14 (\text{\footnotesize lumi})$ \\
    $ZZ$ 8 TeV & $7.1 ^{+0.5}_{-0.4} (\text{\footnotesize stat}) \pm
    0.3 (\text{\footnotesize syst}) \pm 0.2 (\text{\footnotesize
      lumi})$ & $8.4 \pm 1.0 (\text{\footnotesize stat}) \pm
    0.7 (\text{\footnotesize syst}) \pm 0.4 (\text{\footnotesize
      lumi})$ \\ \hline
    $WZ$ 7 TeV & $19.0 ^{+1.4}_{-1.3} (\text{\footnotesize stat}) \pm
    0.9 (\text{\footnotesize syst}) \pm 0.4 (\text{\footnotesize
      lumi})$ & $17.0 \pm 2.4 (\text{\footnotesize stat}) \pm
    1.1 (\text{\footnotesize syst}) \pm 1.0 (\text{\footnotesize
      lumi})$ \\
    $WZ$ 8 TeV & $20.3 ^{+0.8}_{-0.7} (\text{\footnotesize stat})
    ^{+1.2}_{-1.1} (\text{\footnotesize syst}) ^{+0.7}_{-0.6}
    (\text{\footnotesize lumi})$ & \\ \hline
  \end{tabular}
  \it{\caption[]{\small Total cross sections measured by ATLAS and CMS
      experiments at the LHC in $WW$, $ZZ$ and $WZ$ channels (in pb). The
      latter is the sum of $W^+Z$ and $W^- Z$ channels. The references
    for these experimental results can be found in the
    text where the luminosity used is given.}\label{table:vv-exp-results}}
  \end{center}
  \vspace*{-0.1cm}
\end{table}
The experimental results are summarized in
\tab{table:vv-exp-results} and the associated references are
given above with the luminosity used to measure each cross
section. When comparing with our theoretical predictions, the
experimental errors are summed in quadrature to obtain the final
experimental uncertainty. In the case of $WZ$ production ATLAS and CMS
Collaborations give their results on the sum of $W^+Z$ and $W^-Z$
channels. We then add our theoretical predictions together and add in
quadrature the associated uncertainties as we have treated separately
the two channels. This then gives the following results: 
\begin{enumerate}[$\bullet$]
\item{\underline{\it $WW$ channel}: we have at 7 TeV $\sigma^{\rm ATLAS}_{WW} =
    51.9 \pm 4.8$ pb and $\sigma^{\rm CMS}_{WW} = 52.4 \pm 5.1$ pb, giving
    $1.1\sigma$ and $1.2\sigma$ agreements respectively with our
    theoretical prediction $\sigma^{\rm th}_{WW}=45.7^{+3.4}_{-2.6}$ pb. At
    8 TeV, we have $\sigma^{\rm CMS}_{WW} = 69.9 \pm 7.0$ pb, giving a
    $1.8\sigma$ agreement with our theoretical prediction $\sigma^{\rm
      th}_{WW} = 55.6^{+4.0}_{-3.1}$ pb.}
\item{\underline{\it $ZZ$ channel}: we have at 7 TeV $\sigma^{\rm ATLAS}_{ZZ} =
    6.7 ^{+0.9}_{-0.8}$ pb and $\sigma^{\rm CMS}_{ZZ} = 6.24 ^{+0.96}_{-0.87}$
    pb, giving $0.8\sigma$ and $0.3\sigma$ agreements respectively with our
    theoretical prediction $\sigma^{\rm th}_{ZZ}=5.95^{+0.45}_{-0.35}$ pb. At
    8 TeV, we have $\sigma^{\rm ATLAS}_{ZZ} = 7.1^{+0.6}_{-0.5}$ pb and $\sigma^{\rm
      CMS}_{ZZ}= 8.4\pm 1.3$ pb, giving $0.3\sigma$ and $0.8\sigma$
    agreements respectively with our theoretical prediction
    $\sigma^{\rm th}_{ZZ} = 7.3^{+0.5}_{-0.4}$ pb.}
\item{\underline{\it $WZ$ channel}: we have at 7 TeV $\sigma^{\rm ATLAS}_{WZ} =
    19.0^{+1.7}_{-1.6}$ pb and $\sigma^{\rm CMS}_{WZ} =  17.0 \pm 2.8$ pb, giving
    $0.3\sigma$ and $0.4\sigma$ agreements respectively with our
    theoretical prediction $\sigma^{\rm th}_{WZ}=18.3^{+1.2}_{-1.0}$ pb. At
    8 TeV, we have $\sigma^{\rm ATLAS}_{WZ} = 20.3^{+1.6}_{-1.4}$ pb,
    giving a $1.2\sigma$ agreement with our theoretical prediction
    $\sigma^{\rm th}_{WZ} = 22.7^{+1.4}_{-1.1}$ pb.}
\end{enumerate}
We obtain an overall good agreement with the experimental results, in
particular in the $ZZ$ and $WZ$ channels with less than one standard
deviation difference between theory and experiment. This can also be seen
 in \fig{VV-total-error} where the experimental
points together with their uncertainty bands are displayed on top of
our theoretical curves. The highest deviation is seen in the $WW$ channel in
particular at 8 TeV, where the experimental measurement is $1.8\sigma$ higher
than the theoretical prediction.

Given the size of the gluon fusion contribution, which is formally a
NNLO contribution and that one could expect being the dominant
contribution because of the large gluon PDF, we estimate the size of
the missing NNLO contributions in $WW$ production to be of the same
order, namely $2\%$ on top of the full NLO result. We also assume
that this will reduce the scale uncertainty down to $1\%$ at 8 TeV
which gives a $5\%$ total uncertainty. When comparing to the CMS
result, the deviation we get is $1.7\sigma$. This means
that, we think, a full NNLO calculation cannot account for this
deviation. Note that after having written this, the new
paper~\cite{Dawson:2013lya} appeared and its conclusion supports our
estimation.

\section{Conclusion}

We have studied in this paper the full NLO predictions for massive
gauge boson pair production at the LHC, including both QCD and EW
corrections. The latter have been calculated using both mass
regularization and dimensional regularization schemes, including for
the first time the photon-quark induced processes, and our results
are in perfect agreement with each other. Furthermore, our results
without the photon-quark induced processes are similar to the results
of Refs.~\cite{Bierweiler:2012qq, WW-EW-Kuhn, Kuhn-2}, obtained with a
different scale choice. In \sect{section-3} we have presented a study
of the differential distributions and we have analyzed the hierarchy
that is observed in the size of the gluon-quark induced and
photon-quark induced corrections between $ZZ$, $WW$ and $WZ$
channels. Thanks to analytical leading-logarithmic approximations we
have provided the first comprehensive explanation of this hierarchy,
essentially due to the non-abelian gauge structure of the SM,
different coupling strengths and PDF effects. We have found that the
photon-quark induced corrections are negligible in the $ZZ$ channel,
but play an important role in the EW corrections of $WW$ and $WZ$
channels, compensating or even overcompensating the Sudakov virtual
effects. This is because, even though the photon PDF is suppressed, a
new enhancement mechanism with a $t$-channel exchange of a gauge boson
occurs in the hard processes for $WW$ and $WZ$ channels.

In \sect{section-4} we have studied the total cross sections and the
theoretical uncertainties that affect the predictions. The parametric
errors on the $W$ and $Z$ masses have been found negligible, and the
sum of the scale uncertainty and the PDF+$\alpha_s$ uncertainty are
limited to less that $\pm 9\%$ in all channels at all c.m. energy
ranging from 7 to 33 TeV. We have also studied the spread in the
predictions using different PDF sets and have found a $\simeq 4\%$
deviation at maximum with respect to the default MSTW2008 prediction
in all channels. The comparison with ATLAS and CMS experimental
results has also been done and we have found excellent agreement in the
$ZZ$ and $WZ$ channels at 7 and 8 TeV. The agreement in the $WW$
channel is at the $1\sigma$ level at 7 TeV and at the $1.8\sigma$
level at 8 TeV when comparing to the CMS result. We have estimated that
even if a full NNLO result were available this level of agreement
would not significantly differ. This experimental enhancement has
triggered some analyses in terms of new physics to explain it, for
example supersymmetric explanations with charginos
effects~\cite{charginos-WW} or stops contributions~\cite{stops-WW}.

\bigskip

{\bf Acknowledgments:}
We thank Dao Thi Nhung and Dieter Zeppenfeld for fruitful
discussions. This work is supported by the Deutsche
Forschungsgemeinschaft via the Sonderforschungsbereich/Transregio
SFB/TR-9 Computational Particle Physics.

\appendix

\section{Analytical calculation of
  \texorpdfstring{$d\sigma(u\gamma \to W^+ Z d)$}{dsigma(u photon ->
    W+ Z d)} in leading-logarithmic approximation \label{appendixA}}

We present in this appendix the details of the calculation, in leading-logarithmic approximation,
of the EW photon-quark induced process $u \gamma \to W^+ Z d$ in the high $p_T^{Z}$
regime. This process is chosen as an example because it is the most complicated one
including all important features.
The following notations, already introduced in \sect{section-3}, will
be used for the coupling of gauge bosons to quarks,
\beq
c_{L,q} = \frac{1}{\sin\theta_W \cos\theta_W}\left(I_3^q-Q_q
  \sin^2\theta_W\right),\,\, \displaystyle a_W = \frac{1}{\sqrt{2}
  \sin\theta_W}.
\eeq
One has also the following relations between the couplings,
\begin{align}
c_{L,u} - c_{L,d} = \cot\theta_W,\,\, Q_u - Q_d = 1,\,\,
Q_u = a_W^2 - \cot\theta_W c_{L,u},\,\, Q_d = -a_W^2 - \cot\theta_W c_{L,d}.
\label{eq:coupling_relations}
\end{align}

\begin{figure}[h]
  \centering
  \begin{subfigure}[b]{0.45\textwidth}
    \centering
    \includegraphics[scale=0.5]{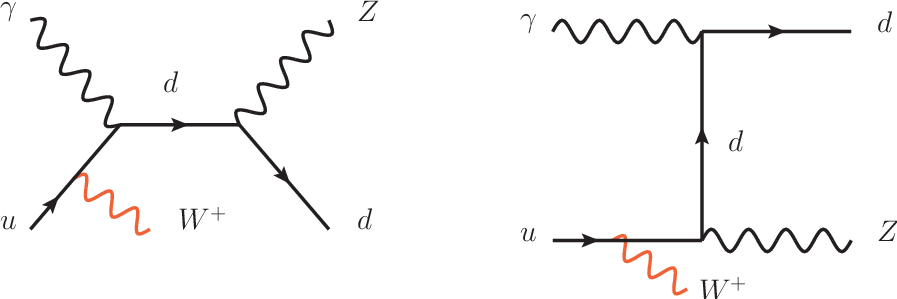}
    \caption{}
    \label{fig:WZ_dgamZd}
  \end{subfigure}
  \hspace{1cm}
  \begin{subfigure}[b]{0.45\textwidth}
    \centering
    \includegraphics[scale=0.5]{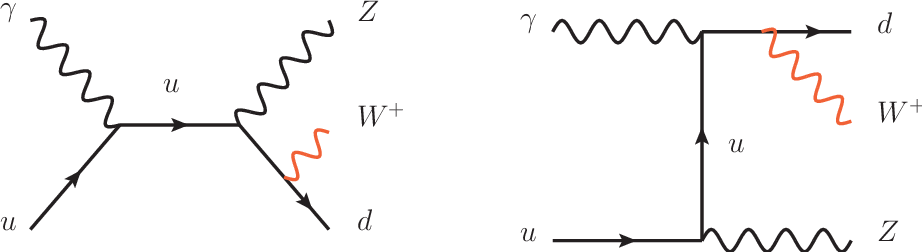}
    \caption{}
    \label{fig:WZ_ugamZu}
  \end{subfigure}\\
  \vspace{3mm}
  \begin{subfigure}[b]{0.45\textwidth}
    \centering
    \includegraphics[scale=0.5]{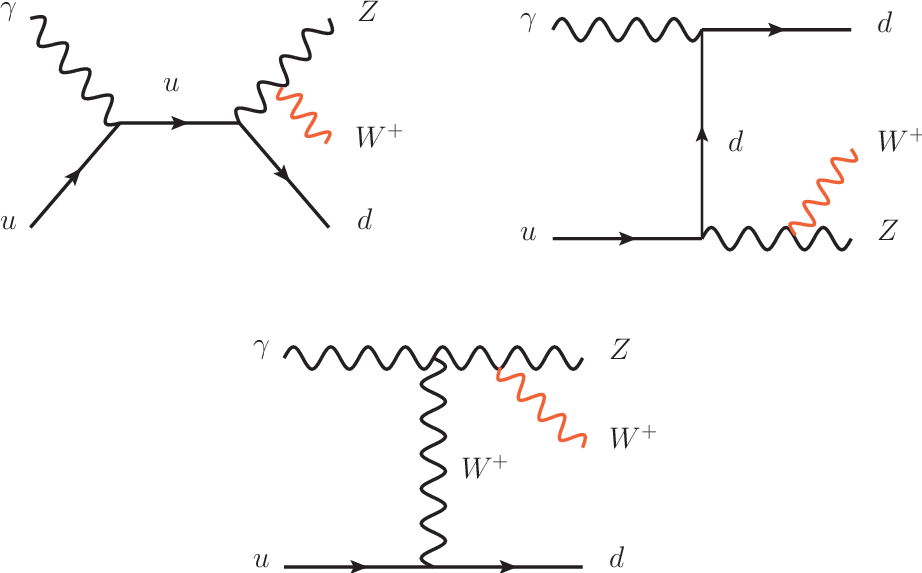}
    \caption{}
    \label{fig:WZ_ugamWd}
  \end{subfigure}
  \hspace{1cm}
  \begin{subfigure}[b]{0.45\textwidth}
    \centering
    \includegraphics[scale=0.5]{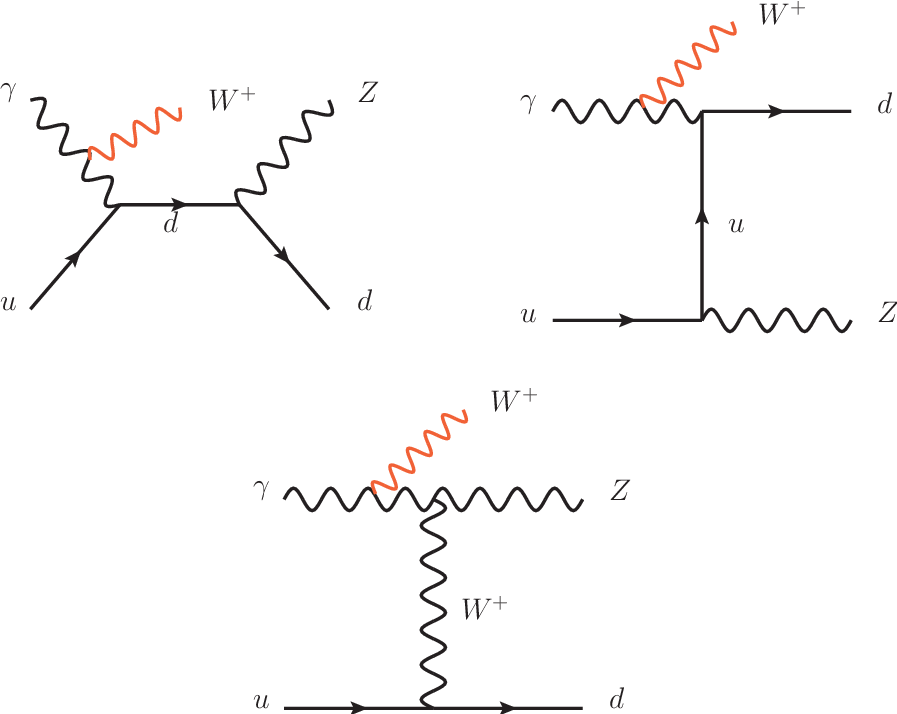}
    \caption{}
    \label{fig:WZ_uWdZ}
  \end{subfigure}
  \caption{\small Diagrams for $u\gamma\to W^+ Z d$ production cross
    section in the case of the differential $p_T^Z$ distribution. The
    soft $W^+$ is displayed in red and the diagrams are classified in
    four categories depending on the $2\to 2$ hard $Z$ subprocess: a)
    is for $d\gamma\to Z d$, b) is for $u\gamma \to Z u$, c) is
    for $u\gamma\to W^+ d$ and d) is for $u W^- \to Z d$.}
  \label{fig:ugamWZd_analytic}
\end{figure}

The leading-logarithmic contribution for the process $u\gamma \to W^+ Z d$
in the high $p_T^Z$ limit is calculated by considering soft $W^+$ radiation.
Thus we display in \fig{fig:ugamWZd_analytic} the Feynman diagrams
for this process with a classification in four categories depending on
the $2\to 2$ production process of the hard $Z$ boson. In \fig{fig:WZ_dgamZd}
one has the soft radiation of the $W^+$ boson from the initial up
quark line, in \fig{fig:WZ_ugamZu} there is a soft radiation from the
final down quark line, in \fig{fig:WZ_ugamWd} the $W^+$ radiated off
the final $Z$ line and in \fig{fig:WZ_uWdZ} the soft radiation is from
the initial photon line.

Before calculating the process, we note that the amplitudes only make
use of left-handed quarks as a $W$ boson is produced. In any $2\to 2$
process involving two gauge bosons we also note that the only helicity
amplitudes surviving at high energy are those where the two gauge
bosons are either both transverse or both
longitudinal. This conclusion agrees with Ref.~\cite{Baur-PRL}.
We then have the following amplitudes,
\beq
  \mathcal{M}_a = e a_W \frac{p_u . \varepsilon^*(k)}{p_u . k}
  \mathcal{A}_L^{d\gamma\to Z d},\, & \displaystyle \mathcal{M}_b = -
  e a_W \frac{p_d . \varepsilon^*(k)}{p_d . k}
  \mathcal{A}_L^{u\gamma\to Z u},\,
  \nonumber\\
  \mathcal{M}_c = e\cot\theta_W \frac{p_Z . \varepsilon^*(k)}{p_Z . k}
  \mathcal{A}_{L}^{u\gamma\to W^+ d},\, & \displaystyle \mathcal{M}_d
  = - e \frac{p_\gamma . \varepsilon^*(k)}{p_\gamma . k}
  \mathcal{A}_{LT}^{u W^- \to Z d},\,
  \label{eq:amplitude}
\eeq
where $k$ stands for the momentum of the soft $W^+$ boson, the
subscript $L$ means that all quarks are left-handed and $T$ means that
all gauge bosons are transverse. As we work in the high $p_T$ regime
we have $p^2 \to 0$ for any of the external particle (in the
leading-logarithmic approximation). This means that only the
interference terms between the various amplitudes $\mathcal{M}_i$
survive, then killing the longitudinal term in $\mathcal{M}_d$ as we
always have a photon involved in the interference between
$\mathcal{M}_d$ and the other amplitudes. This explains why we have
already discarded this longitudinal term in \eq{eq:amplitude} above,
the other amplitudes being always transverse at high energy as they
involve a photon. This argument that only transverse gauge bosons can
contribute to the leading-logarithmic results also holds for the $ZZ$
and $W^+W^-$ cases and is crucial for obtaining simple results.

When integrating over the $W^+$ boson phase-space we use the following
soft approximation,
\beq
\int_{M_W}^E \frac{d^3 k}{(2\pi)^3 2 \omega_k}  \frac{2 p_i p_j}{(p_i
  . k)(p_j . k)} =
\frac{1}{8\pi^2}\log^2\left(\frac{p_T^2}{M_W^2}\right)\ \text{ in the
  limit } M_W^2 \ll p_T^2 \simeq E^2\, .
\label{eq:softintegral}
\eeq
When squaring the sum of the amplitudes in \eq{eq:amplitude} and using
\eq{eq:softintegral} one has
\begin{align}
\vert \bar{\mathcal{A}}^{u \gamma\to W^+ Z d} \vert^2 =  \displaystyle \frac12 \Big[ &
a_W^2 \Re\left(\mathcal{A}_L^{d\gamma\to Z d}
  {\mathcal{A}_L^{u\gamma\to Z u}}^*\right) + \cot\theta_W \Re\left(
  \mathcal{A}_{L}^{u\gamma\to W^+ d} {\mathcal{A}_{LT}^{u W^-\to Z
      d}}^*\right)+ \nonumber\\
 &  a_W \Re\left( {\mathcal{A}_L^{d\gamma\to Z d}}^* ( -\cot\theta_W
   \mathcal{A}_{L}^{u\gamma\to W^+ d} + \mathcal{A}_{LT}^{u W^-\to Z
     d}) \right) - \nonumber\\
 & a_W \Re\left( {\mathcal{A}_L^{u\gamma\to Z u}}^* ( -\cot\theta_W
   \mathcal{A}_{L}^{u\gamma\to W^+ d} + \mathcal{A}_{LT}^{u W^-\to Z
     d}) \right) \Big] \frac{\alpha}{2\pi}
\log^2\left[\frac{(p_T^Z)^2}{M_W^2}\right]\, ,
\label{eq:cross_section}
\end{align}
where it is implicitly assumed that the sum over color is done
in the real part of the amplitude products. The quarks are
taken left-handed only in the $2\to 2$ subprocesses which implies the
extra one-half factor coming from the spin average in the unpolarized
$2\to 3$ process. The gauge bosons are all transverse but yet
unpolarized and their spin average is implicitly assumed in the above
expression.

The key point is now to relate the two amplitudes
$\mathcal{A}_{L}^{u\gamma\to W^+ d}$ and $\mathcal{A}_{LT}^{u W^-\to
  Z d}$ to the amplitude $\mathcal{A}_L^{u\gamma\to Z u}$. By using
only transverse gauge bosons one can link the amplitude
${\mathcal{A}_3}_{L}^{u\gamma\to W^+ d}$ displayed by the third
diagram of \fig{fig:WZ_ugamWd} to the amplitude
${\mathcal{A}_3}_{LT}^{u W^-\to Z d}$ displayed by the third diagram
of \fig{fig:WZ_uWdZ} for the $2\to 2$ hard subprocess: $\displaystyle
{\mathcal{A}_3}_{LT}^{u W^-\to Z d} = \cot\theta_W
{\mathcal{A}_3}_{L}^{u\gamma\to W^+ d}$. This gives
\beq
\mathcal{A}_{L}^{u\gamma\to W^+ d} & = & \frac{a_W}{c_{L,u}}
{\mathcal{A}_1}_L^{u\gamma\to Z u} + \frac{Q_d}{Q_u}
\frac{a_W}{c_{L,u}} {\mathcal{A}_2}_L^{u\gamma\to Z u}  +
{\mathcal{A}_3}_{L}^{u\gamma\to W^+ d}\, ,\nonumber\\
\mathcal{A}_{LT}^{u W^-\to Z d} & = & -\frac{a_W}{Q_u}
\frac{c_{L,d}}{c_{L,u}} {\mathcal{A}_1}_L^{u\gamma\to Z u}  -
\frac{a_W}{Q_u} {\mathcal{A}_2}_L^{u\gamma\to Z u} + \cot\theta_W
{\mathcal{A}_3}_{L}^{u\gamma\to W^+ d}\, ,
\eeq
from which one obtains the following equation, with the use of
\eq{eq:coupling_relations},
\beq
\cot\theta_W \mathcal{A}_{L}^{u\gamma\to W^+ d} - \mathcal{A}_{LT}^{u
  W^-\to Z d} = a_W a_u \mathcal{A}_L^{u\gamma\to Z u},
\label{eq:key_relation}
\eeq
with $a_u = 1 - (Q_d c_{L,d})/(Q_u c_{L,u})$ as defined in \eq{eq:au_ad}.
This (which is the same as \eq{eq_amp_wz}) is the master equation to
express the differential cross section in a simple form. We also use
the following equation
\beq
\mathcal{A}_{L}^{d\gamma\to Z d} & = &  \frac{Q_d}{Q_u}
\frac{c_{L,d}}{c_{L,u}} \mathcal{A}_{L}^{u\gamma\to Z u}
\eeq
to replace the $d\gamma\to Z d$ subprocess by the $u\gamma\to Z u$
subprocess. Using \eq{eq:key_relation} two times one can link not only
the amplitudes together but also the squared amplitudes, giving
{\small \begin{align}
2\cot\theta_W \Re\left(\mathcal{A}_{L}^{u\gamma\to W^+ d} {\mathcal{A}_{LT}^{u
    W^-\to Z d}}^*\right) = \cot^2\theta_W \left|\mathcal{A}_{L}^{u\gamma\to W^+
  d}\right|^2 + \left|\mathcal{A}_{LT}^{u W^-\to Z d}\right|^2 - a_W^2
a_u^2 \left|\mathcal{A}_{L}^{u\gamma\to Z u}\right|^2 \! .
\label{eq:key_relation2}
\end{align}}
The final result is obtained when inserting \eq{eq:key_relation} and
\eq{eq:key_relation2} into \eq{eq:cross_section}, leading to
{\small \begin{align}
d\sigma^{u \gamma\to W^+ Z d}  = \left[
  \fr{a_W^2}{2}\left(1-a_u+\frac{a_u^2}{2}\right)d\sigma_L^{u\gamma\to
    Z u}
+ \frac{\cot^2\theta_W}{4}
d\sigma_{L}^{u\gamma\to W^+ d} + \frac{1}{4} d\sigma_{LT}^{u W^- \to Zd}
\right] \frac{\alpha}{2\pi}
\log^2\left[\frac{(p_T^Z)^2}{M_W^2}\right],
\label{eq:WpZ_ew_final}
\end{align}}
where the photon PDF has to be used, at the hadronic level, for the
initial $W^-$. This ends our proof of \eq{eq_log2_wz_EW_softW} where
the result for $d \gamma\to W^- Z u$ process is also given. Similar
results for $q \gamma\to Z Z q$, $q \gamma\to W^\pm Z q^\prime$ and $q
\gamma\to W^+ W^- q$, obtained using the above explained method, can
be found in \sect{section-ZZ}, \sect{section-WZ} and
\sect{section-WW}, respectively. The result for the QCD gluon-quark induced
process $ug \to W^+Zd$ can be derived from \eq{eq:WpZ_ew_final} using
the following rules: $\gamma \to g$, $Q_q \to 1$,
$d\sigma_{L}^{u\gamma \to W^+ d} \to d\sigma_{L}^{ug \to W^+ d}$ and
$d\sigma_{LT}^{u W^- \to Zd} \to 0$.

\clearpage

\bibliography{main}
\bibliographystyle{h-physrev}

\end{document}